\documentclass[aps, pre, onecolumn, nofootinbib, notitlepage, groupedaddress, amsfonts, amssymb, amsmath, longbibliography]{revtex4-1}
\usepackage{tabularx}
\usepackage{graphicx}
\usepackage{hyperref}
\usepackage{xcolor}
\hypersetup{
    colorlinks,
    linkcolor={red!50!black},
    citecolor={blue!50!black},
    urlcolor={blue!80!black}
}
\usepackage{bm}
\usepackage{natbib}
\usepackage{longtable}
\newcommand{\DivU}{\ensuremath{\nabla\cdot\bm{u}}}
\newcommand{\angles}[1]{\ensuremath{\left\langle #1 \right\rangle}}
\newcommand{\KS}[1]{\ensuremath{D_{\text{KS}}(#1)}}
\newcommand{\KSstat}[1]{\ensuremath{\overline{D_\text{KS}(#1)}}}
\newcommand{\grad}{\ensuremath{\nabla}}
\newcommand{\RB}{Rayleigh-B\'{e}nard }

\newcommand{\approptoinn}[2]{\mathrel{\vcenter{
	\offinterlineskip\halign{\hfil$##$\cr
	#1\propto\cr\noalign{\kern2pt}#1\sim\cr\noalign{\kern-2pt}}}}}

\begin{document}
\author{Evan H. Anders}
\affiliation{Dept. Astrophysical \& Planetary Sciences, University of Colorado -- Boulder, Boulder, CO 80309, USA}
\affiliation{Laboratory for Atmospheric and Space Physics, Boulder, CO 80303, USA}
\author{Benjamin P. Brown}
\affiliation{Dept. Astrophysical \& Planetary Sciences, University of Colorado -- Boulder, Boulder, CO 80309, USA}
\affiliation{Laboratory for Atmospheric and Space Physics, Boulder, CO 80303, USA}
\author{Jeffrey S. Oishi}
\affiliation{Department of Physics and Astronomy, Bates College, Lewiston, ME 04240, USA}
\title{Accelerated evolution of convective simulations}

\begin{abstract}
High Peclet number, turbulent convection is a classic system with a large timescale
separation between flow speeds and the thermal relaxation time.
In this paper, we present a method of
fast-forwarding through the long thermal relaxation of convective simulations, and we test the
validity of this method. This Accelerated Evolution (AE) method involves measuring the dynamics of convection
early in a simulation and using its characteristics to adjust the mean thermodynamic
profile within the domain towards its evolved state. We study \RB convection as a test case for AE.  
Evolved flow properties of AE solutions are measured to be within a few percent
of solutions which are reached through Standard Evolution (SE) over a full thermal diffusion timescale.
At the highest values of the Rayleigh number at which we compare SE and AE,
we find that AE solutions require roughly an order of magnitude fewer computing hours
to evolve than SE solutions.
\end{abstract}
\maketitle


\section{Introduction}
\label{sec:intro}
Astrophysical convection occurs in the presence of disparate timescales. 
Studying realistic models of natural systems through direct numerical
simulations is infeasible because of the
large separation between various flow timescales and relaxation times.
Stiffness in astrophysical systems can manifest in multiple ways, some of which can
be handled by clever choices of numerical algorithms, and some which cannot.
For example,
flows in the convection zones of stars like the Sun are characteristically low Mach number
(Ma) in the deep interior. Initial value problems solved using
explicit timestepping methods are bound by the Courant-Friedrich-Lewy
(CFL) timestep limit corresponding to the fastest motions (sound
waves), resulting in timesteps which are prohibitively
small for studies of the deep, low-Ma motions. These systems are numerically
stiff, and the difference between
the sound crossing time and the convective overturn time has made studies of low-Ma stellar
convection difficult. This stiffness can be avoided using approximations such as
the anelastic approximation, in which sound waves are explicitly filtered out
\cite{brown&all2010, featherstone&hindman2016}.
Recently, advanced numerical techniques which use fully implicit 
\cite{viallet&all2011, viallet&all2013, viallet&all2016} or mixed
implicit-explicit \cite{lecoanet&all2014, anders&brown2017, bordwell&all2018} 
timestepping mechanisms have made it possible to study
convection in the fully compressible equations at low Mach numbers, 
and careful studies of deep convection which
would have been impossible a decade ago are now widely accessible.

Unfortunately, astrophysical convective systems are stiff in more than one 
timescale. Specifically, the Peclet number (Pe), the ratio of the thermal
diffusion timescale to the convective velocity timescale, is large.
In a high Pe system, many convective timescales must pass before 
the domain thermally relaxes into a steady state in which
evolution of the thermal structure of the convective region has ceased.
As the timestep size of implicit methods is bound to the fastest nonlinear flows
(convection), fully implicit methods cannot be used to address this form of stiffness
\cite{viallet&all2011, viallet&all2013, viallet&all2016}. 

Resolving dynamics in atmospheres which are sufficiently
thermally relaxed therefore remains a challenging problem.
Solar convection is a prime example of this phenomenon, as
dynamical timescales in the solar convective zone are relatively short 
(convection overturns every $\sim 5$ min at the solar surface)
compared to the Sun's thermal relaxation timescale, which is O($10^7$) years
\cite{stix2003}.  
In such a system, it is impossible to resolve the convective dynamics while also
meaningfully evolving the thermal structure of the system using
traditional timestepping techniques alone.
As modern simulations aim to model natural, high-Pe convection
in the high-Rayleigh-number (Ra) regime,
the thermal diffusion timescale ($t_{\kappa}$, defined in Sec. \ref{sec:ae}) 
becomes intractably large compared to dynamical timescales 
such as the freefall time ($t_{\text{ff}}$, defined in Sec. \ref{sec:experiment})
\cite{anders&brown2017}, 
\begin{equation}
\frac{t_{\kappa}}{t_{\text{ff}}} \propto (\text{Ra})^{1/2}.
\end{equation}
Furthermore, as dynamical and thermal timescales separate, 
simulations become more turbulent. Capturing appropriately resolved
turbulent motions requires finer grid meshes and smaller timesteps.
Thus, the progression of simulations into the high-Ra
regime of natural convection is slowed by two simultaneous effects: timestepping
through a single convective overturn time becomes more computationally expensive
and the number of overturn times required for systems to thermally relax
grows.

The vast difference between convective and thermal timescales has long plagued
numericists studying convection, and an abundance of approaches has been employed to
study thermally relaxed solutions. One popular method for accelerating the convergence
of high-Ra solutions is by ``bootstrapping'' -- the process of using the relaxed flow
fields and thermal structure of a low Ra solution as initial conditions for a simulation at high
Ra.  This method has been used with great success \cite{johnston&doering2009, verzicco&camussi1997},
but it is not without its faults.  In systems in which there are multiple
stable solutions, such as a roll state and a shearing state of convection,
the large-scale convective structures present in the
low Ra solution imprint onto the dynamics of the new, high Ra solution. 
It is possible that this puts the high Ra solution into a different stable state
than it would naturally reach from an initial, hydrostatically stable configuration.
Another commonly-used tactic in
moderate-Ra simulations is to first solve a simple model of the system in question,
and then use the solution of that model as initial conditions for full nonlinear
direct numerical simulations. For example, past studies have used initial conditions from
a linear eigenvalue solve \cite{hurlburt&all1984} in plane-parallel studies, or
axisymmetric solutions in studies of convection in three-dimensional (3D)
 cylinders \cite{verzicco&camussi1997}. 
In other systems, particularly when convective zones are adjacent to stable regions,
authors often choose initial conditions which are not in a classic, hydrostatic,
conductive state. Rather, either through knowledge of low-Ra solutions \cite{couston&all2017}
or broader convective theories such as Mixing Length Theory
\cite{brandenburg&all2005}, initial conditions can be adjusted such that the stratification
within the convective domain is closer to a relaxed state than the largely
unstable hydrostatic state.

Despite the numerous methods that have been used,
the most straightforward way to achieve a thermally relaxed solution
is to evolve a convective simulation through a thermal timescale. Some modern
studies do just that \cite{featherstone&hindman2016}.
Such evolution is computationally 
\emph{expensive}, and state-of-the-art simulations at the highest values of Ra
can only reasonably be run
for hundreds of freefall timescales \cite{stevens&all2011}, much less the
thousands or millions of freefall times required for thermal relaxation.

In this work, we study a method of achieving accelerated evolution of
convective simulations. 
Our technique is similar in essence to the approach used in asymptotically 
reduced models of rapidly rotating convection, in which the
mean temperature profile evolves separately from the
fast convective dynamics \cite{julien&all1998, sprague&all2006}. 
We couple measurements of the dynamics of unstable, evolving
convective simulations with knowledge about energy balances in the relaxed solution
to instantaneously and self-consistently adjust the mean vertical thermodynamic 
profile toward its relaxed state. 
While a technique of this kind has been used by many studies previously 
(e.g., \cite{hurlburt&all1986}),
the details of implementation, the convergence properties, and whether or not the
thermally relaxed state achieved from 
accelerated evolution corresponds to the relaxed state in standard evolution 
are not documented.
In Sec. \ref{sec:experiment}, we describe our convective simulations and
numerical methods. In Sec. \ref{sec:ae}, we describe the procedure for 
achieving accelerated evolution. In
Sec. \ref{sec:results}, we compare accelerated evolution solutions
to solutions obtained from standard evolution through a full thermal diffusion timescale. 
In Sec. \ref{sec:speedups}, we compare the numerical cost of accelerated
solutions to standard solutions.
Finally,
in Sec. \ref{sec:extensions}, we offer concluding remarks and
discuss extensions of the methods presented here.


\section{Experiment}
\label{sec:experiment}
In this work we study a simple form of thermal convection:
incompressible \RB convection under the Oberbeck-Boussinesq approximation,
such that the fluid
has a constant kinematic viscosity ($\nu$), thermal diffusivity ($\kappa$), and coefficient
of thermal expansion ($\alpha$). The density of the fluid is a constant, $\rho_0$,
except in the buoyancy term, where it is $\rho = \rho_0(1  - \alpha T_1)$.
The gravitational acceleration, $\bm{g} = - g\bm{\hat{z}}$, is constant.
The equations of motion are \cite{spiegel&veronis1960}:
\begin{gather}
\DivU = 0, 
	\label{eqn:incompressible}
\\
\frac{\partial \bm{u}}{\partial t} + \bm{u}\cdot\grad\bm{u} =
-\frac{1}{\rho_0}\grad P - g( 1 - \alpha T_1)\hat{z} + \nu\grad^2\bm{u}, 
	\label{eqn:dim_bouss_momentum}
\\
\frac{\partial T_1}{\partial t} + \bm{u}\cdot\grad(T_0 + T_1) = \kappa\grad^2 (T_0 + T_1),
	\label{eqn:dim_bouss_energy}
\end{gather}
where $\bm{u} = u\bm{\hat{x}} + v\bm{\hat{y}} + w\bm{\hat{z}}$ is the velocity, 
$T = T_0(z) + T_1(x, y, z, t)$ are the initial and fluctuating components of temperature, 
and $P$ is the kinematic pressure. The initial temperature profile, $T_0$, decreases
linearly with height.
We non-dimensionalize these equations such that
length is in units of the layer height ($L_z$),
temperature is in units of the initial temperature jump across the layer ($\Delta T_0 = L_z \partial_z T_0$), 
and velocity is in units of the freefall velocity ($v_{\text{ff}} = \sqrt{\alpha g L_z^2 \partial_z T_0}$).
By these choices, one time unit is a freefall time ($t_{\text{ff}} = L_z/v_{\text{ff}}$).
We introduce a reduced kinematic pressure,
$\varpi \equiv (P / \rho_0 + \phi + |\bm{u}|^2 / 2) / v_{\text{ff}}^2$, where the gravitational
potential, $\phi$, is defined such that $\bm{g} = -\grad \phi$. 
In non-dimensional form, and substituting 
$\bm{u}\cdot\grad\bm{u} = \grad(|\bm{u}|^2/2) - \bm{u}\times(\grad\times\bm{u})$
and $\grad^2\bm{u} = -\grad\times(\grad\times\bm{u})$, Eqs.~(\ref{eqn:dim_bouss_momentum}) and (\ref{eqn:dim_bouss_energy})
become
\begin{align}
\frac{\partial \bm{u}}{\partial t} + \grad \varpi - T_1\hat{z} + \mathcal{R}\grad\times\bm{\omega} &= \bm{u}\times\bm{\omega},
	\label{eqn:bouss_momentum}
\\
\frac{\partial T_1}{\partial t} - \mathcal{P}\grad^2 T_1 + w \frac{\partial T_0}{\partial z} &= - \bm{u}\cdot\grad T_1,
	\label{eqn:bouss_energy}
\end{align}
where $\bm{\omega} = \grad \times \bm{u}$ is the vorticity.
The dimensionless control parameters $\mathcal{R}$ and $\mathcal{P}$ 
are set by the Rayleigh (Ra) and Prandtl (Pr) numbers,
\begin{equation}
\mathcal{R} \equiv \sqrt{\frac{\text{Pr}}{\text{Ra}}}, \qquad \mathcal{P} \equiv \frac{1}{\sqrt{\text{Pr}\,\text{Ra}}}, \qquad
\text{Ra} = \frac{g \alpha L_z^4 \partial_z T_0}{\nu\kappa} = \frac{(L_z\,v_{\text{ff}})^2}{\nu\kappa}, \qquad \text{Pr} = \frac{\nu}{\kappa}.
\end{equation}
We hold Pr $= 1$ constant throughout this work, such that $\mathcal{P} = \mathcal{R}$.
$\mathcal{P}$ and $\mathcal{R}$ are related to the inverse Reynolds and Peclet numbers of the
evolved flows.

In Eqs.~(\ref{eqn:incompressible}), (\ref{eqn:bouss_momentum}), and (\ref{eqn:bouss_energy}),
linear terms are grouped on the left-hand side of the equations, while nonlinear terms
are found on the right-hand side. We timestep linear terms implicitly, and nonlinear
terms explicitly.
We utilize the 
Dedalus\footnote{\url{http://dedalus-project.org/}} 
pseudospectral framework \cite{burns&all2016} to evolve  
Eqs.~(\ref{eqn:incompressible}), (\ref{eqn:bouss_momentum}), and (\ref{eqn:bouss_energy}) 
forward in time
using an implicit-explicit (IMEX), third-order, four-stage 
Runge-Kutta timestepping scheme RK443 \cite{ascher&all1997}. The code used to run the simulations
in this work is included in the supplemental materials as a zip file \cite{supp}.

Variables are time-evolved on a dealiased Chebyshev (vertical)
and Fourier (horizontal, periodic) domain in which the
physical grid dimensions are 3/2 the size of the coefficient grid.  
We study two- (2D) and three-dimensional (3D) convection in which the domain is a cartesian box, 
whose dimensionless vertical extent is $z \in [0, 1]$, 
and which is horizontally periodic with an extent of $x, y \in [0, \Gamma]$,
where $\Gamma = 2$ is the aspect ratio, as has been previously studied
\cite{goluskin&all2014, johnston&doering2009}. 
In 2D simulations, we set $v = \partial_y = 0$.
We specify no-slip, impenetrable boundary conditions at both the top and
bottom boundaries,
\begin{equation}
u = v = w = 0 \, \, \text{at}\,\,z = 0,1.
\label{eqn:vel_bcs}
\end{equation}
The temperature is fixed at the top boundary, and the flux is fixed at the
bottom boundary, such that
\begin{equation}
T_1 = 0 \,\,\text{at}\,\, z=1, \qquad
\frac{\partial T_1}{\partial z} = 0\,\,\text{at}\,\,z=0.
\label{eqn:temp_bcs}
\end{equation}
For this choice of boundary conditions, the critical value of Ra at which
the onset of convection occurs is Ra$_{\text{crit}} = 1295.78$ \cite{goluskin2016}, and the
supercriticality of a run is defined as $S \equiv \text{Ra}/\text{Ra}_{\text{crit}}$.
Studies of convection which aim to model
astrophysical systems such as stars often employ mixed thermal
boundary conditions, as we do here \cite{hurlburt&all1984, cattaneo&all1991, korre&all2017}.
Our choice of the thermal boundary conditions in Eqn.~(\ref{eqn:temp_bcs}) 
was motivated by the fact that accelerated evolution is simpler when both the
thermal profile and the flux through the domain are fixed at a boundary 
(see Sec. \ref{sec:ae}).

The initial temperature profile is linearly unstable,
$T_0(z) = 0.5 - z$. On top of this profile, we fill $T_1$ with
random white noise whose magnitude is $10^{-6}\mathcal{P}$, and which is
vertically tapered so as to match the thermal boundary conditions.
This ensures that the initial perturbations are much smaller than the
evolved convective temperature perturbations, even at large Ra.
We filter this noise spectrum in coefficient space, 
such that only the lower 25\% of the coefficients
have power; this low-pass filter is used to avoid populating the
highest wavenumbers with noise in order to improve the stability of our
spectral timestepping methods.


\section{The method of Accelerated Evolution}
\label{sec:ae}
Here we describe a method of Accelerated Evolution (AE), which we use 
to rapidly relax the thermodynamic state of convective simulations.  
We compare this AE method to Standard Evolution
(SE), in which we evolve the atmosphere from noise initial conditions
for one thermal diffusion time,
$t_\kappa = \mathcal{P}^{-1}$. Both AE and SE simulations begin with identical
initial conditions, as described in Sec. \ref{sec:experiment}.
As Ra increases, and $\mathcal{P}$ decreases, SE solutions become intractable, 
while the timeframe of convergence for an AE solution remains nearly constant
in simulation freefall time units (see table \ref{table:run_parameters} in
appendix \ref{appendix:run_table}).

\begin{figure}[p!]
\includegraphics[width=\textwidth]{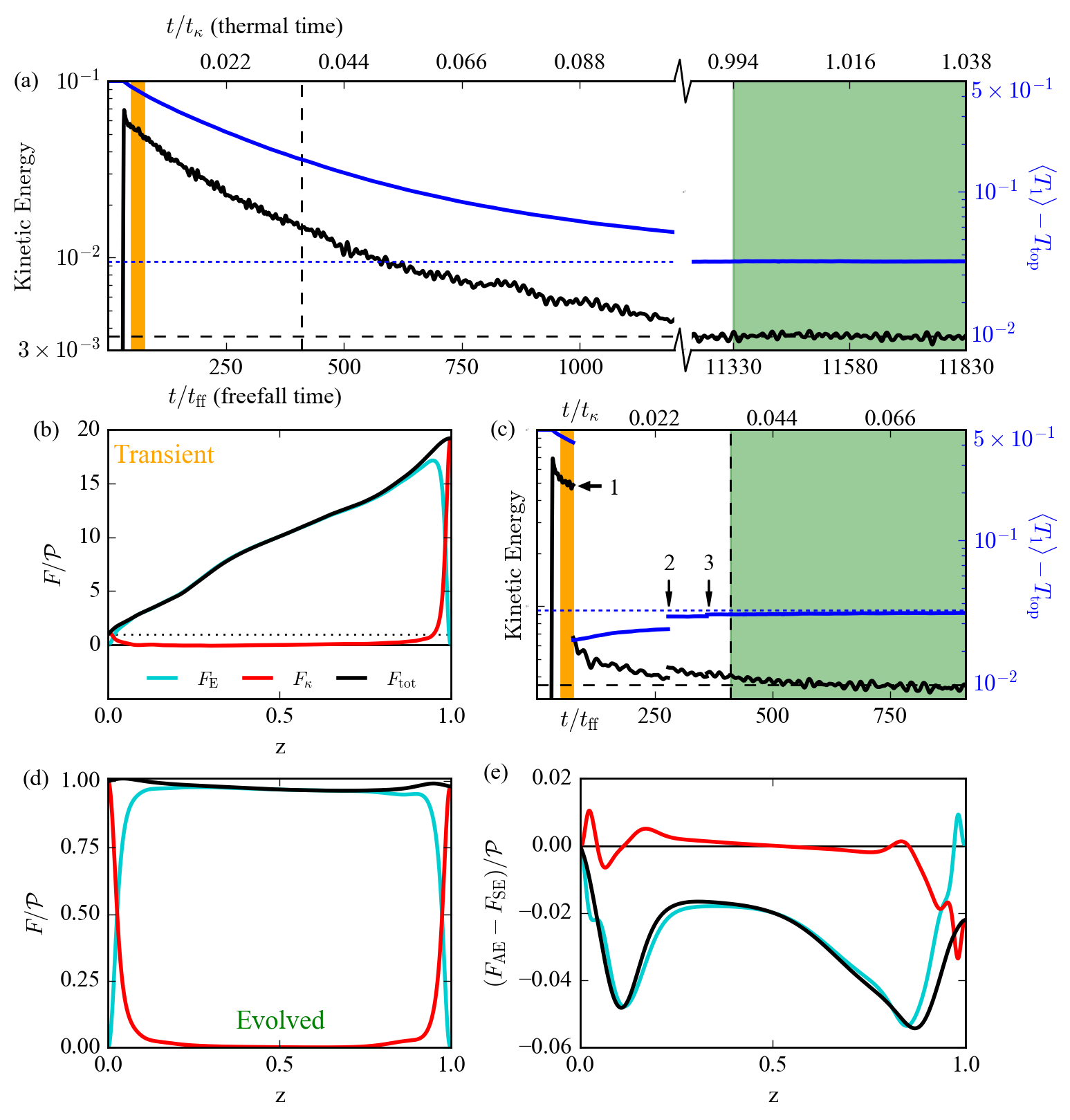}
\caption{(a) Kinetic energy (black) and $\angles{T} - T_{\text{top}}$ (blue)  vs. time are shown
for a SE run at $S = 10^5$. The mean evolved values of kinetic energy and mean temperature,
averaged over the time shaded in green,
are denoted by the horizontal dashed lines. (b) The time- and horizontally-averaged
flux profiles are shown for the times highlighted in orange in (a).
(c) The same quantities as in (a) are shown, but for AE at the same parameters.
The axes are scaled identically in (a) and (c), and the AE method is used three times, marked by
the numbered arrows. The fluxes averaged over the green shaded region of (c)
are shown in (d). The difference between
the fluxes in the AE and SE solutions is shown in (e). \label{fig:time_trace} }
\end{figure}

We study in depth a 2D simulation at $S = 10^5$ to demonstrate the power of AE.
We compare kinetic energy (KE, black line) and mean temperature (blue line)
traces from a SE run in Fig.~\ref{fig:time_trace}(a) to an AE run
in Fig.~\ref{fig:time_trace}(c).
In Fig.~\ref{fig:time_trace}(a), the time evolution of the SE simulation is shown.
The KE grows exponentially from white noise during the
first $\sim 25$ $t_{\text{ff}}$. The solution then saturates and begins to slowly
relax toward the saturated isothermal profile in the interior of the domain.
This slow relaxation is evident in the behavior of the blue line, which measures
$\angles{T} - T_{\text{top}}$, where \angles{T} is the volume-average of $T$, and
$T_{\text{top}} = -0.5$ is the temperature at the upper boundary.
The mean atmospheric temperature and
kinetic energies are fully converged when $t = 4000t_{\text{ff}} = 0.35t_{\kappa}$.
We show roughly the first thousand freefall time
units of evolution, as well as the evolved thermodynamic state reached after a full
thermal time of evolution.  In Fig.~\ref{fig:time_trace}(c), similar traces are
shown for the corresponding AE solution
at the same parameters. The same linear growth phase occurs, but shortly after
the peak of convective transient we accelerate the convergence of the atmosphere
through the process which we describe below. We adjust the 1D vertical profile of
the atmosphere three times, as denoted by the three labeled
arrows in the graph numbered 1-3.  The third profile adjustment associated with
arrow 3 is small enough (see appendix \ref{appendix:recipe}) that we assume
the atmosphere is sufficiently converged, and we begin to sample the evolved
convective dynamics.

The horizontally averaged profiles of the vertical conductive flux, 
F$_{\kappa} = \angles{-\kappa\partial_z T}_{x,y}$, and the vertical convective enthalpy flux,
F$_{\text{E}} = \angles{wT}_{x,y}$, are the basis of the AE method.
Here we use $\angles{}_{x,y}$ to represent a horizontal average. We measure
both of these fluxes early in a simulation, retrieving profiles such as
those shown in Fig.~\ref{fig:time_trace}(b).
As the atmosphere relaxes towards
the isothermal profile specified by the upper (cold) boundary condition, excess
temperature throughout the atmosphere must leave the domain. This excess thermal
energy leaves through the upper boundary, as seen in 
Fig.~\ref{fig:time_trace}(b), where the amount of flux exiting at the top of the domain
is nearly 20 times larger the flux entering the bottom of the domain. Once the atmospheric
temperature profile reaches its evolved state, the flux entering the bottom boundary
is equal to the flux exiting through the upper boundary.  In general, this 
evolution is slow in SE [Fig.~\ref{fig:time_trace}(a)], but AE [Fig.~\ref{fig:time_trace}(c)]
can rapidly advance a system whose fluxes are in a strongly disequilibrium state [Fig.~\ref{fig:time_trace}(b)],
into a near-equilibrium state, as shown in Fig.~\ref{fig:time_trace}(d). In
this final state both boundaries conduct the same amount of flux.
The converged fluxes achieved through AE are at most 5\% different from the SE solution, 
as shown in Fig.~\ref{fig:time_trace}(e). 
This is a very small difference considering the short timescales on which convergence
is reached and the strongly disequilibrium state used to inform the AE process.

In order to adjust the temperature profile to achieve AE, we calculate the total flux,
F$_{\text{tot}} =$ F$_{\text{E}}$ + F$_{\kappa}$, and then derive the profiles
\vspace{-0.5cm}
\begin{equation}
f_{\text{E}}(z) = \frac{\text{F}_{\text{E}}}{\text{F}_{\text{tot}}},\qquad
f_{\kappa}(z) = \frac{\text{F}_{\kappa}}{\text{F}_{\text{tot}}},
\label{eqn:bvp_ratios}
\end{equation}
which have the systematic asymmetries [Fig.~\ref{fig:time_trace}(b)] removed. These profiles describe which
parts of the atmosphere depend on convection to carry flux (where $f_{\text{E}}(z) = 1$
and $f_{\kappa}(z) = 0$).
We presume that the early convection occupies roughly the same volume as the evolved
convection, and thus that the extent of the early thermal boundary layers 
(where $f_{\kappa}(z) = 1$ and $f_{\text{E}}(z) = 0$) 
will not change significantly over the course of the atmosphere's evolution.
Under this assumption, in order to reach the converged state, 
the flux through the atmosphere must be decreased by some amount,
\vspace{-0.5cm}
\begin{equation}
\xi(z,t) \equiv \frac{\text{F}_{\text{B}}}{\text{F}_{\text{tot}}},
\label{eqn:xi}
\end{equation}
where $\text{F}_{\text{B}} = \mathcal{P}$ is the amount of flux that enters the
atmosphere at the bottom, fixed-flux boundary.
For example, in Fig.~\ref{fig:time_trace}b,
F$_{\text{tot}} \approx 19\mathcal{P}$ at the upper boundary,
but in the relaxed state 
it should just be $\mathcal{P}$, so $\xi \approx 1/19$ at that depth.

To reduce the conductive flux by $\xi$, we examine the 
horizontally- and time-averaged
Eqs.~(\ref{eqn:bouss_momentum}) and (\ref{eqn:bouss_energy}) in the time-stationary state; after 
neglecting terms which vanish due to symmetry, these equations become
\vspace{-0.2cm}
\begin{gather}
\frac{\partial}{\partial z}\angles{\varpi}_{x,y} - \angles{T_1}_{x,y}\hat{z} = \angles{\bm{u}\times\bm{\omega}}_{x,y, \text{ ev}},
	\label{eqn:bouss_BVP_momentum}
\\
\frac{\partial}{\partial z}\text{F}_{\text{E, ev}} - \mathcal{P}\frac{\partial^2}{\partial z^2} \angles{T_1}_{x,y} = 0.
	\label{eqn:bouss_BVP_energy}
\end{gather}
Here, we construct  $\text{F}_{\text{E, ev}} = \xi \text{F}_{\text{E}}$ 
and $\angles{\bm{u}\times\bm{\omega}}_{x, y,\text{ ev}}
= \xi\angles{\bm{u}\times\bm{\omega}}_{x, y}$.
Using these profiles and 
Eqs. (\ref{eqn:bouss_BVP_momentum}) and (\ref{eqn:bouss_BVP_energy}),
we solve for $\angles{\varpi}_{x,y}$ and $\angles{T_1}_{x,y}$.
Our choice of fixing the temperature at the top boundary and the flux at the bottom
boundary ensures that there is a unique solution for $\angles{T_1}_{x,y}$ given the evolved
fluxes. If flux were fixed at both boundaries, there would be infinitely many
temperature solutions. If temperature were fixed at both boundaries, 
$\text{F}_{\text{B}}$ would not be precisely known \emph{a priori}.

Solving the above equations adjusts the mean thermal state and 
$\text{F}_{\kappa}$ of the atmosphere while leaving $\text{F}_{\text{E}}$
unchanged. In the  
bulk of the domain, convective enthalpy flux dominates transport, and so there we
assume that $\text{F}_{\text{tot}} \approx \text{F}_{\text{E}}$. Upon 
removing terms which vanish due to symmetry, we note that the convective
enthalpy flux is carried solely by the velocity field and perturbations in
temperature away from the mean state, $\text{F}_{\text{E}} = 
\angles{w(T_1 - \angles{T_1})}$.
In order to instantaneously decrease the convective enthalpy flux in a manner
which is consistent with the conductive flux, 
we multiply both the velocity, $\bm{u}$,
and temperature perturbations about the mean state, $T_1-\angles{T_1}_{x,y}$,
 by $\sqrt{\xi}$. This scaling of the velocity field is the reason that we
multiply $\angles{\bm{u}\times\bm{\omega}}_{x, y}$, which is nondimensionally
of order $u^2/L$, by $\xi$ while constructing
$\angles{\bm{u}\times\bm{\omega}}_{x, y,\text{ ev}}$.

In general, the AE method occurs in the following steps. First, we pause a convective simulation and
construct $\xi(z,t)$ from measured flux profiles in the convective domain.
We solve a 1D boundary value problem (BVP) consisting of
Eqs.~(\ref{eqn:bouss_BVP_momentum}) and (\ref{eqn:bouss_BVP_energy})
to obtain an evolved thermodynamic profile, and its corresponding conductive flux.
We multiply both the temperature perturbations around the mean and the
convective velocity flows by $\sqrt{\xi}$.
After adjusing the fields of a simulation in this manner, we continue timestepping forward.
For specifics on the precise implementation of the AE method, we refer
the reader to appendix \ref{appendix:recipe}.

\section{Results}
\label{sec:results}

\begin{figure}[b]
\includegraphics[width=\textwidth]{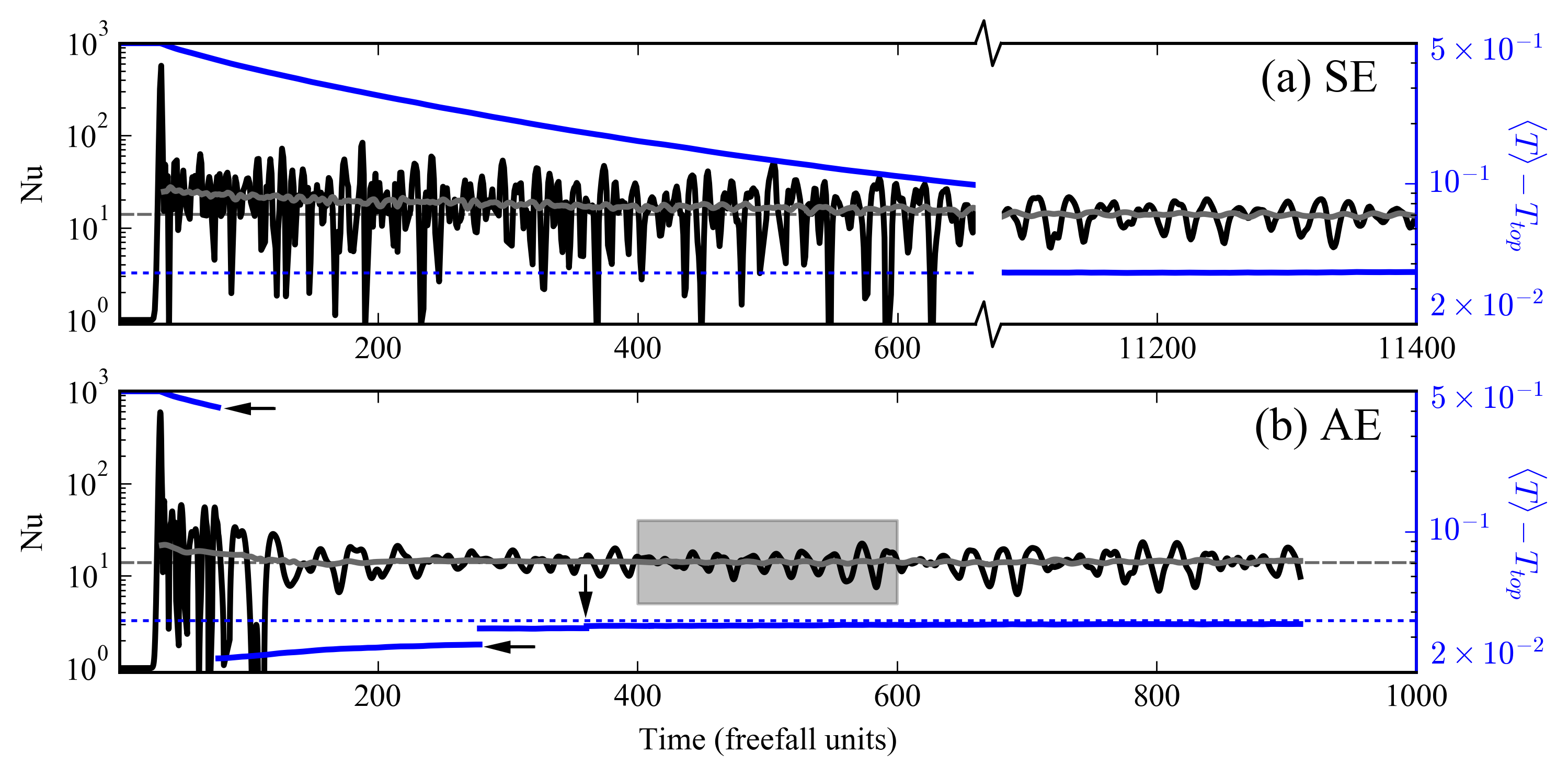}
\caption{The time evolution of Nu (black) and $\angles{T} - T_{\text{top}}$ (blue)
are shown for simulations at $S = 10^5$ in SE (a) and AE (b).
A moving average of Nu, using a centered boxcar with a span of 50 freefall time
units, is overplotted as a gray line. 
We see that the mean value of Nu reaches its relaxed value quickly
compared to $\angles{T}$.
Regardless, fluctuations of Nu about the mean
value are much smaller in the thermally relaxed state.
The times at which AE solves occur are denoted by arrows in (b),
and the gray boxed region is examined in more detail in Fig. \ref{fig:oscillating_plumes}.
\label{fig:nu_v_time} }
\end{figure}

We study evolved standard evolution (SE) solutions whose supercriticalities ($S$) are 
$S \in (1, 10^5]$ in 2D and $S \in (1, 10^4]$ in
3D. We compare their properties to
accelerated evolution (AE) runs at $S \in (1, 10^7]$ in 2D and
$S \in (1, 10^4]$ in 3D.
We refer the reader to appendix \ref{appendix:run_table} for a full list of
simulations.

The Nusselt number (Nu) quantifies the efficiency of convective heat transport
and is defined as
\begin{equation}
\text{Nu} = \frac{\angles{F_{\text{conv}} + F_{\text{cond}}}}{\angles{F_{\text{cond, ref}}}}
 = \frac{\angles{wT - \mathcal{P}\partial_z T}}{\angles{- \mathcal{P} \partial_z T}},
\end{equation}
where the volume average of a quantity, $\eta$, is shown as $\angles{\eta}$.
The time evolution of Nu in AE and SE is compared to the mean temperature evolution
in Fig.~\ref{fig:nu_v_time}. In Fig.~\ref{fig:nu_v_time}(a), we show the evolution
of the SE run at $S = 10^5$. The black trace shows the instantaneous value of
Nu, and the overplotted gray line shows a moving time average of Nu. The
time average is taken using a centered boxcar window whose width is 50 $t_{\text{ff}}$.
The mean value of Nu evolves towards its relaxed value (dashed horizontal
gray line) rapidly compared to $\angles{T}-T_{\text{top}}$ (blue), 
which approaches its evolved state (dashed horizontal blue line) slowly.
As such, it is possible to measure the statistically stationary mean value of
Nu after only a few hundred simulated freefall times, 
as has been done previously \cite{stevens&all2010}. However, fluctuations in
Nu around the mean value at early times have both a larger magnitude and higher
frequency than the fluctuations in the relaxed state.
In Fig.~\ref{fig:nu_v_time}(b), we show the AE traces of Nu and
$\angles{T}-T_{\text{top}}$ at $S = 10^5$. The times at which AE solves occur are marked
by arrows. Nu reaches its evolved mean value slightly more rapidly than in the SE case,
and the frequency and magnitude of fluctuations in Nu away from the mean resemble the
final relaxed state of SE.

\begin{figure}[t]
\includegraphics[width=\textwidth]{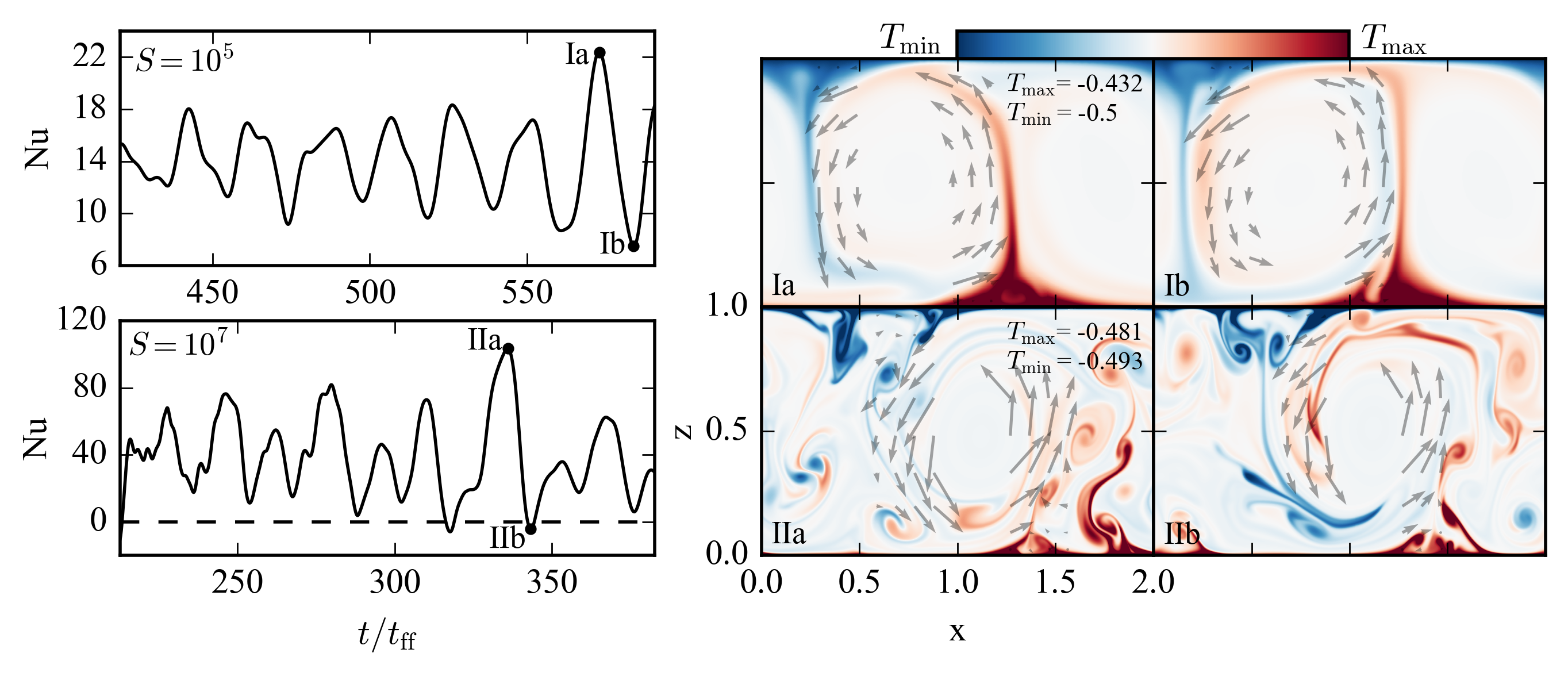}
\caption{The time variation of the Nusselt number is shown for two AE cases at
$S = 10^5$ (top) and $S = 10^7$ (bottom). On the left, the instantaneous value of Nu
is shown as a function of time. On the right, temperature snapshots are shown for
Nu maxima (Ia and IIa) and minima (Ib and IIb). The suppressed value of Nu at the
minma arises from entrainment of fluid elements whose temperature perturbations
are wrongly signed (e.g., hot material going downwards and cold material going
upwards in Ib and IIb). The colorbar scaling of panels Ia\&b are the same, as
are the scalings of panels IIa\&b. 
The minimum temperature is chosen by the fixed-temperature
boundary condition at the top, $T_{\text{top}} = -0.5$. The decreased range of
the colorbar scaling for
for IIa\&b was chosen to better display the convective dynamics.
\label{fig:oscillating_plumes} }
\end{figure}

When $S < 10^{3.67}$ in 2D and for all runs in 3D, 
the evolved system is characterized by a time-stationary value of Nu, and is thus
in a state of constant convective heat transport.
At larger $S$ in 2D, the value of Nu varies significantly over time even in the
relaxed state (as seen in Fig.~\ref{fig:nu_v_time}). We examine the shaded
region of Fig.~\ref{fig:nu_v_time}(b) in more detail in the top left
panel of Fig.~\ref{fig:oscillating_plumes}, as well as a comparable time trace
at $S = 10^7$ (bottom, left panel). We find that these systems exhibit
large Nu during
states in which temperature fluctuations travel in their natural buoyant
direction (Fig.~\ref{fig:oscillating_plumes}, Ia and IIa, where cold elements fall and hot elements rise).
However, when wrongly-signed temperature perturbations are entrained in an upflow or downflow
with oppositely signed fluid, Nu is suppressed (Fig.~\ref{fig:oscillating_plumes}, Ib and IIb, 
where warm fluid is pulled down by the downflow
lane, and cool fluid is drawn up by the upflow lane).
The plumes in these
systems naturally oscillate horizontally over time, switching between transport being dominated
by a counterclockwise cell, as pictured in Fig.~\ref{fig:oscillating_plumes}, and
a clockwise cell. Our choice of no-slip
boundary conditions prevents the fluid from entering a full domain shearing state 
\cite{goluskin&all2014}, and the
oscillatory motions of the plumes are a long-lived, stable phenomenon. 
However, thanks to our choice of periodic
boundary conditions and despite the no-slip conditions, the full system of the
upflow and downflow plumes is free to slowly migrate to the left or right over time,
and we observe this phenomenon in our solutions.
The 2D SE simulations exhibit the same horizontally
oscillatory behavior as the AE solutions for the same initial conditions. 
This time-dependent behavior of Nu is not seen strongly in our 3D solutions,
however most 3D simulations we conducted were at low $S$ compared to the runs in
which this behavior was observed in 2D.

\begin{figure}[t]
\vspace{-1.5cm}
\includegraphics[width=\textwidth]{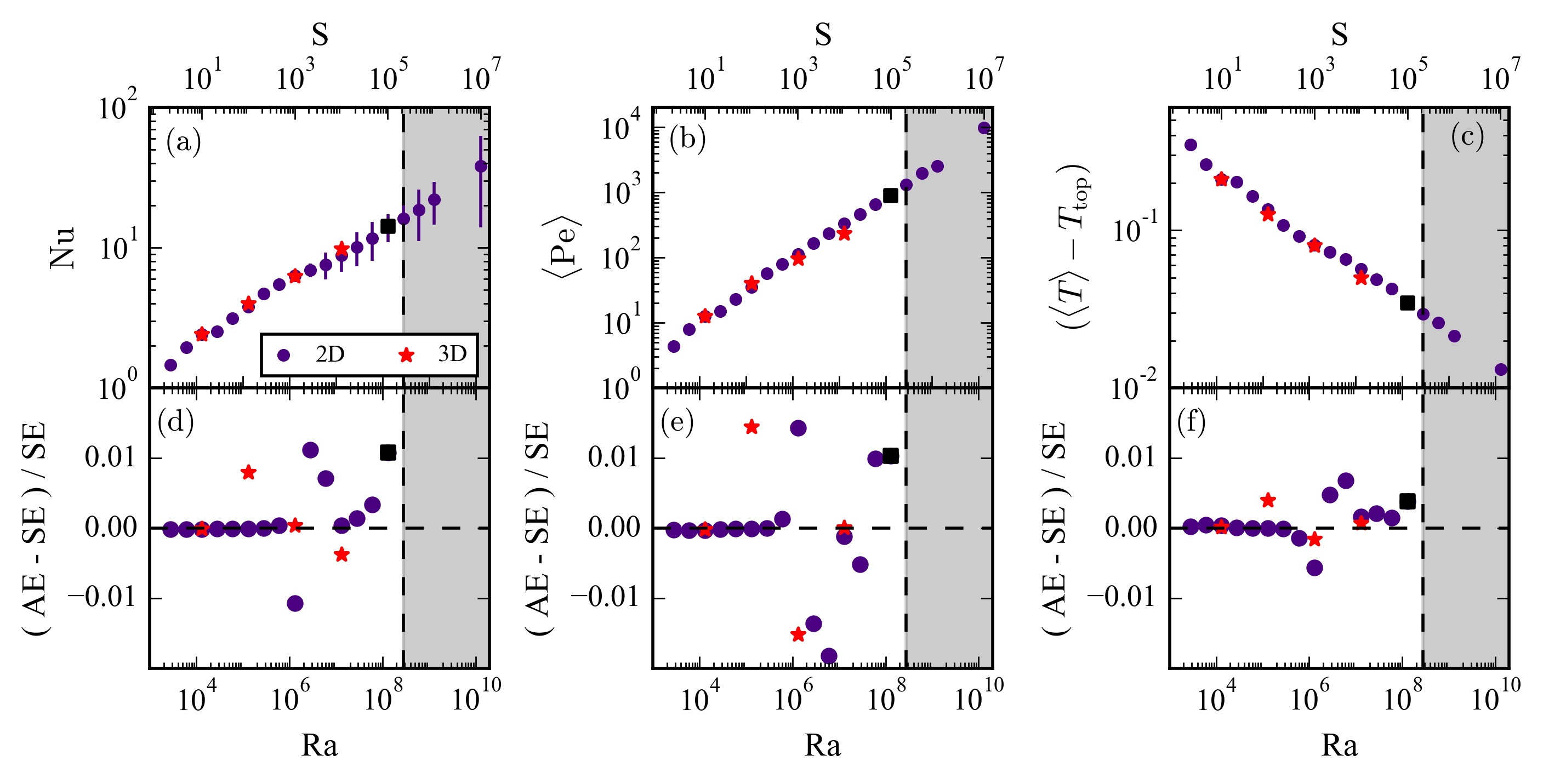}
\caption{Volume- and time-averaged measurements of the Nusselt number (Nu), the
RMS Peclet number ($\angles{\text{Pe}}$), and the mean temperature ($\angles{T}$) for AE runs are shown in (a)-(c).
Symbols are located at the mean value of
each measurement and denote 2D (purple circles) and 3D (red stars). 
The run at $S = 10^5$ marked as a
black square is examined in more detail in Figs. \ref{fig:time_trace}-\ref{fig:oscillating_plumes},
\ref{fig:temp_comparison}, and \ref{fig:pdf_comparison}.
Vertical lines represent the standard deviation of the measurement,
and quantify natural variation over the averaging window. 
(a) Nu scales as Ra$^{1/5}$; at high $S$ in 2D the value of Nu fluctuates over time
(see Fig.~\ref{fig:oscillating_plumes}).  
(b) Pe, which measures turbulence in the solution, scales as
Ra$^{0.45}$. (c) The difference between $\angles{T}$ and the value of $T$ at the fixed-temperature
top boundary is shown; this quantity scales as Ra$^{-1/5}$, the inverse of Nu.
Relative error for measurements of (d) Nu, (e) Pe, and (f) $\angles{T} - T_{\text{top}}$ between 
AE solutions and SE solutions are shown.
The greyed area of the plots indicates the region in which only AE runs were
carried out due to computational expense. \label{fig:parameter_space_comparison} }
\end{figure}

The time- and volume-averaged values of Nu, the RMS
Peclet number (Pe), and the mean temperature 
are shown for AE solutions in Figs.~\ref{fig:parameter_space_comparison}(a)-\ref{fig:parameter_space_comparison}(c).
Mean values are shown by the symbols (purple circles and red stars), and 
the vertical lines represent the standard deviation of the measurement over time.
Nu is shown as a function of Ra and $S$ in 
Fig.~\ref{fig:parameter_space_comparison}a, while 
$\text{Pe} = \angles{|\bm{u}|} / \mathcal{P}$ is shown in 
Fig.~\ref{fig:parameter_space_comparison}(b), and $\angles{T} - T_{\text{top}}$ 
(the mean temperature value minus its value at the upper, 
fixed temperature boundary) is shown in Fig.~\ref{fig:parameter_space_comparison}(c).
We report $\text{Nu} \propto \text{Ra}^{1/5}$,
$\text{Pe} \propto \text{Ra}^{0.45}$, and 
$(\angles{T} - T_{\text{top}}) \propto \text{Ra}^{-1/5}$.
The average temperature, $\angles{T}$, 
is dominated by its value in the isothermal interior,
so this measurement serves as a probe of the temperature jump across the boundary
layers. Thus, the inverse scaling of average temperature and Nu that we
find here is expected for thermally equilibrated solutions \cite{otero&all2002}.

In Figs.~\ref{fig:parameter_space_comparison}(d)-\ref{fig:parameter_space_comparison}(f), we report the fractional difference
between measurements in the AE and SE solutions.
The mean values of Nu and $\angles{T} - T_{\text{top}}$ 
measured in AE are accurate to SE values to within $\sim 1$\%.
Pe measurements show marginally greater error, with AE measurements being 
$\leq 2$\% different from SE measurements.

For the select 3D runs conducted in this study, the scaling of Nu, Pe, and $\angles{T} - T_{\text{top}}$
reported in Figs.~\ref{fig:parameter_space_comparison}(a)-\ref{fig:parameter_space_comparison}(c) is nearly identical to the
2D simulations. Errors between AE and SE solutions in 3D fall within the same range as
errors in 2D in Figs.~\ref{fig:parameter_space_comparison}(d)-\ref{fig:parameter_space_comparison}(f). AE is therefore
equally effective in both 2D and 3D, and we restrict much of our study to 2D here
in order to more thoroughly sample parameter space.

The measurements presented in Fig.~\ref{fig:parameter_space_comparison} demonstrate
that AE can be powerfully employed in parameter space studies in which
large numbers of simulations are compared in a volume-averaged sense.  We now turn
our examination to a more direct comparison of AE and SE for 2D convection at
$S = 10^5$, the time, flux, and Nu evolution of which are shown in 
Figs.~\ref{fig:time_trace}, \ref{fig:nu_v_time}, and \ref{fig:oscillating_plumes}.
All comparisons that follow for these two runs occur over the times shaded in
green in Figs.~\ref{fig:time_trace}(a) and \ref{fig:time_trace}(c). Measurements are sampled every
0.1 freefall time units for 500 total freefall time units.

\begin{figure}[t!]
\includegraphics[width=\textwidth]{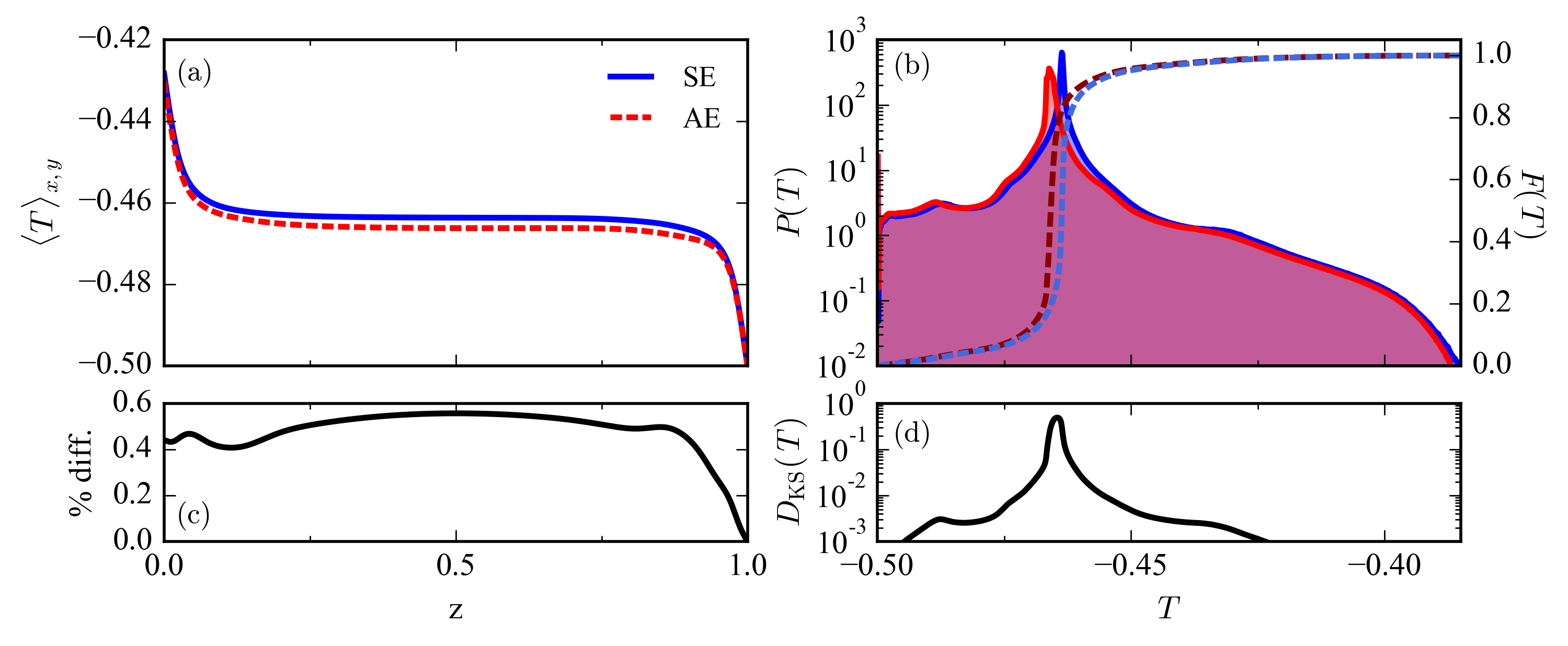}
\caption{Comparisons of the evolved thermodynamic states of an AE and SE run
at $S = 10^{5}$ are shown.  (a) Evolved horizontally- and time-averaged 
temperature profiles, as a function of height.
(b) Probability Distribution Functions (PDFs, $P(T)$) and their integrated
Cumulative Distribution Functions (CDFs, $F(T)$)
of point-by-point measurements of the temperature field.
(c) The percentage difference between the mean temperature profiles as a function of height.
The difference between the mean profiles is very small, O(0.5\%).
(d) $\KS{T}$, as defined in Eqn.~\ref{eqn:ks_profile}, is shown. The small
difference in the mean interior temperature between AE and SE
results in a large difference between the two temperature distributions near the values
of the temperature maxima.  The spread of temperature around the maxima, which includes the 
fluctuations that drive convection, are nearly identical between the two runs.
\label{fig:temp_comparison} }
\end{figure}

As AE is fundamentally a 1D adjustment to the thermodynamic structure of the
solution, we compare the horizontally- and time-averaged temperature profiles 
attained by AE and SE in Fig.~\ref{fig:temp_comparison}(a).  
The boundary layer width and structure are  
nearly identical between the two solutions, but the
the mean temperature in the isothermal interior differs by roughly 0.5\%
[Fig.~\ref{fig:temp_comparison}(c)]. 

The probability distribution functions (PDFs)
of point-by-point temperature measurements are compared for the two runs
in Fig.~\ref{fig:temp_comparison}(b). To construct these PDFs, 
we interpolate the full temperature field
at each measurement time onto an evenly spaced grid, determine the
frequency distribution of all $T$ values over the duration of the 500 $t_{\text{ff}}$
measurement window, and then normalize the
distribution such that its integral is unity.  The two PDFs have noticeably
different maxima, as is expected from Fig.~\ref{fig:temp_comparison}(a). 
Over long timescales, the 0.5\% difference between the two profiles would
disappear, as the AE solution evolves to be exactly the SE solution; this
is evident in the asymmetry of the AE PDF near the maxima
in Fig.~\ref{fig:temp_comparison}(b) and also
the trend of the mean temperature over time in Fig.~\ref{fig:time_trace}(c).

One method of comparing two
PDFs to determine if they are drawn from the same underlying
sample distribution is through the use of a Kolmogorov-Smirnov ($D_{\text{KS}}$) test \cite{wall&jenkins2012}.
We calculate the $D_{\text{KS}}$ statistic for a PDF of some value, $q$, as
\begin{equation}
\KS{q} = F_{\text{AE}}(q) - F_{\text{SE}}(q),
\label{eqn:ks_profile}
\end{equation}
where $F$ stands for cumulative distribution function (CDF), the integral of the PDF.
A traditional Kolmogorov-Smirnov statistic is just a single value,
$\KSstat{q} = |\KS{q}|_\infty =
\text{max} |\KS{q}|$, and we use both the profile KS$(q)$ and
$\KSstat{q}$ to gain insight into the likeness of two PDFs. 
We show $\KS{T}$ in Fig.~\ref{fig:temp_comparison}(d), and the
CDFs used to construct it overlay the PDFs in Fig.~\ref{fig:temp_comparison}(b).
Near the maxima of the temperature PDFs, $\KSstat{T} = 0.495$, 
which is very large and implies that roughly half of all
measurements in the AE case are at a lower $T$ than those in the SE case.
While this difference is significant, it is also expected from Fig.~\ref{fig:temp_comparison}(a).
Fortunately, $\KS{T}$ is very small away from the maxima, 
indicating that the temperature fluctuations off of the maxima, which are the primary
drivers of convective transport, are nearly identical.

\begin{figure}[b!]
\includegraphics[width=\textwidth]{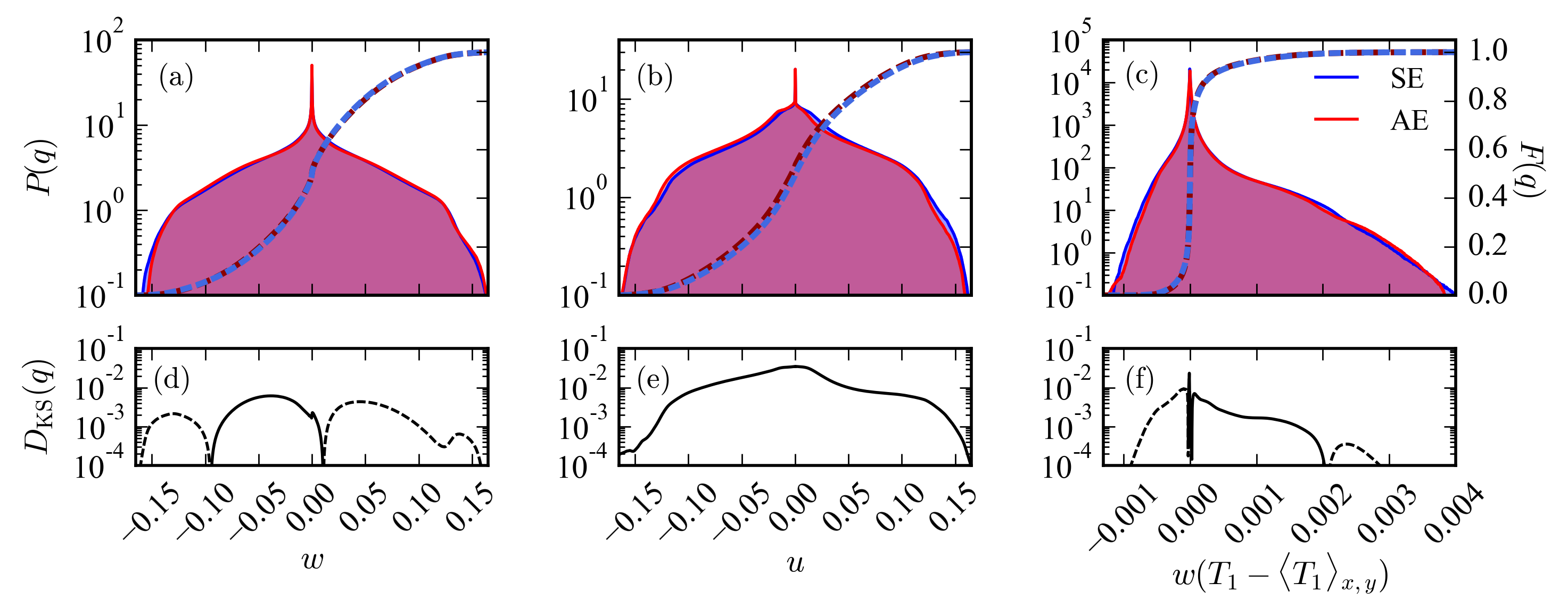}
\caption{Probability distribution functions (PDFs, $P(q)$) of (a) the vertical velocity ($q = w$), (b) the horizontal velocity ($q = u$), and (c) nonlinear
convective transport [$q = w(T_1 - \angles{T_1}_{x,y})$] are shown for 2D runs achieved through SE (blue) and AE (red)
at $S = 10^{5}$.  The cumulative distribution function (CDF) is overplotted for each PDF. 
(d-f) The $D_{\text{KS}}$ profiles, as defined in Eq.~(\ref{eqn:ks_profile}),
are shown for the related distributions; solid lines indicate positive values
while dashed lines are negative values. Unlike the temperature distributions in
Fig.~\ref{fig:temp_comparison}, these distributions
show very good agreement and small values of the $D_{\text{KS}}$ statistic.
\label{fig:pdf_comparison} }
\end{figure}

\newpage$\,$\newpage

In addition to adjusting the 1D thermal profile, the AE method also scales the
simulation velocities and temperature fluctuations by $\sqrt{\xi}$. 
In Fig.~\ref{fig:pdf_comparison}
we examine the velocities and heat transport found in the evolved states.
Shown are the PDFs of 
vertical velocity [$w$, Fig.~\ref{fig:pdf_comparison}(a)], horizontal velocity [$u$, Fig. \ref{fig:pdf_comparison}(b)],
and the nonlinear vertical convective flux [$w(T_1 - \angles{T_1}_{x,y})$, Fig.~\ref{fig:pdf_comparison}(c)]. 
Each PDF here shows a strong peak near zero due to the no-slip, impenetrable
velocity boundary conditions (Eq.~(\ref{eqn:vel_bcs})).
The CDFs of each profile are overplotted, and corresponding KS profiles are
shown in Figs.~\ref{fig:pdf_comparison}(d)-\ref{fig:pdf_comparison}(f).  We report
$\KSstat{w} = 0.00615$, $\KSstat{u} = 0.0349$,
and $\KSstat{w(T_1 - \angles{T_1}_{x,y})} = 0.0263$.
The difference in vertical velocity
and heat transport between AE and SE is negligible, which is unsurprising in light of
the Nu measurements of Figs.~\ref{fig:parameter_space_comparison}(a) and \ref{fig:parameter_space_comparison}(d).
This also confirms that the large $\KSstat{T}$ in Fig.~\ref{fig:temp_comparison}(d) is
not of concern, and that the AE run achieves the same relaxed convective solution
as the SE run.
We find that the difference in $\KS{u}$, which consistently has more probability
of flows moving left (in the -$x$ direction), appears to be caused by a more prominent migration 
of the full roll system in the -$x$ direction in the AE run than in the SE run. 
This migration does not appear to affect the vertical transport appreciably.

\section{Computational Time-savings of AE}
\label{sec:speedups}

\begin{table}[b!]
\caption{Shown are details regarding the computational cost of select
AE and SE runs in 2D and 3D. The supercriticality
($S$), coefficient resolution (nz, nx, ny), number of CPUs used
to perform the calculation ($N_{\text{CPU}}$), number of CPU-hours
used to perform the run ($t_{\text{CPU,AE/SE}}$), and ratio of CPU-hours
used in the AE run compared to the SE run are provided.
}
\setlength{\tabcolsep}{12pt}
\label{table:speed}
\begin{center}
\begin{tabularx}{0.72\textwidth}{ c r r c c c }
\hline																	
$S$	&	nz$\times$nx$\times$ny	&	$N_{\text{CPU}}$	&	
$t_{\text{CPU, SE}}$ & $t_{\text{CPU, AE}}$ &$t_{\text{CPU,AE}}/t_{\text{CPU,SE}}$\\
\hline \hline \multicolumn{6}{c}{\vspace{-0.2cm}}\\
\multicolumn{6}{c}{\vspace{0.1cm}2D Runs} \\
\hline
$10^2$	&	64$\times$128	&	32    &  2.2              & 4.4               &     2.0  \\
$10^3$	&	128$\times$256	&	64    &  53               & 21                &     0.39 \\
$10^4$	&	256$\times$512	&	128   &  1.2$\times 10^3$ & 1.8$\times 10^2$  &     0.15  \\
$10^5$	&	512$\times$1024&	256   &  2.4$\times 10^4$ & 2.8$\times 10^3$  &     0.12 \\
\hline \hline \multicolumn{6}{c}{\vspace{-0.2cm}}\\
\multicolumn{6}{c}{\vspace{0.1cm}3D Runs} \\
\hline
$10^1$	&	32$\times$64$\times$64	    & 512       &   62                 &        1.1$\times 10^2$   & 1.7\\
$10^2$	&	64$\times$128$\times$128	& 512       &   1.9$\times 10^2$   &        1.1$\times 10^2$   & 0.60 \\
$10^3$	&	128$\times$256$\times$256	& 2048      &   7.0$\times 10^3$   &        1.4$\times 10^3$   & 0.20 \\
$10^4$	&	256$\times$512$\times$512	& 8192      &   3.3$\times 10^5$   &        2.3$\times 10^4$   & 0.070\\
\hline																	
\end{tabularx}
\end{center}
\end{table}

Computational time-saving is the primary reason to use AE rather than 
evolving all solutions through SE.
In table 
\ref{table:speed}, we compare cpu-hour cost for select
2D and 3D runs in which both AE and SE solutions were computed. Times reported
for AE and SE runs only include the time required to reach a relaxed
state, and do not include the time over which measurements were taken
in that state (e.g., the green shaded regions of
Figs.~\ref{fig:time_trace}(a) and \ref{fig:time_trace}(c) are not included in the $S = 10^5$ times).
All simulations
were performed on Broadwell nodes on NASA's Pleiades supercomputer 
(Intel Xeon E5-2680v4 processors). The key metric which highlights the usefulness
of AE is the number of cpu-hours used for the AE run divided by cpu-hours used
for the SE run ($t_{\text{CPU,AE}}/t_{\text{CPU,SE}}$). We see that at low
resolution and low supercriticality, $t_{\text{CPU,AE}}/t_{\text{CPU,SE}} > 1$,
and AE is not useful. However, as $S$ grows, $t_{\text{CPU,AE}}/t_{\text{CPU,SE}}$
shrinks. At the highest supercriticalities for which AE and SE were compared
in this work, AE runs cost roughly an order of magnitude less computing time 
than SE runs.

Integrating information about the mean state in time (fluxes, etc.)
decreases the rate at which our solver timesteps early in the AE cases. 
However, 
the first application of AE in a given simulation 
[e.g., Fig.~\ref{fig:time_trace}(c), at the arrow labeled ``1''] drastically
increases the average timestep by fastforwarding the simulation into
a more stable state with lower convective velocities. 
For the $S = 10^5$ case we examined in detail, the
average time step grew by a factor of 2-3.
At $S = 10^7$, the AE solve immediately improved the timestep size 
by nearly a factor of 4.


\newpage
\section{Discussion \& Conclusions}
\label{sec:extensions}
In this work we have studied a method of Accelerated Evolution (AE) which can
be employed to achieve rapid thermal relaxation of convective simulations.  We compared
this technique to the Standard Evolution (SE) of convection through a full thermal diffusion timescale,
and we
showed that AE rapidly obtains solutions whose dynamics are statistically similar to SE solutions.
The AE method is valid at low values of $S$, where SE solutions
converge quickly due to the short thermal timescale, and AE remains applicable
at high values of $S$, where SE solutions are intractable.
As discussed, AE is equally applicable in 2D and 3D; here we have restricted most of our study to 2D
to extend our parameter space coverage.
At the largest values of $S$ in which AE and SE are compared in this work, we find
time savings of nearly an order of magnitude. 

Here we studied the simplest possible case for the application of AE:
\RB convection at low aspect ratio with mixed thermal boundary conditions. 
We anticipate that
this technique will be powerful in its extensions to more complicated studies.
To achieve AE in more complicated systems, one need only derive 
the steady-state, horizontally-averaged equations governing
the convective dynamics
[e.g., the analogs to Eqs.~(\ref{eqn:bouss_BVP_momentum}) and (\ref{eqn:bouss_BVP_energy})]
and couple those equations with knowledge of the boundary conditions
and current dynamics as described in
Sec. \ref{sec:ae} and appendix \ref{appendix:recipe}.
In general, AE should be useful in studies where there are two disparate
timescales which must both be resolved and which cannot be overcome through
clever timestepping techniques.  Some avenues in which extensions of AE could
be beneficial for expanding the available parameter space of exploration
include studies of internally heated convection \cite{goluskin2016},
convection with height-dependent conductivities \cite{kapyla&all2017},
penetrative convection \cite{hurlburt&all1986,brandenburg&all2005,couston&all2017},
or fully compressible, stratified convection \cite{anders&brown2017}.
As AE is fundamentally a horizontally uniform adjustment to the thermodynamic
structure of the convective domain, it is unlikely that these techniques
should be applied straightforwardly to nonperiodic convective domains.

We conclude by noting that AE should be extended to these more complicated
studies with caution. 
While AE was extremely effective in this simple case studied here
(where the aspect ratio was low, 
the bounds of the convective domain were pre-defined,
and the solutions were simple rolls),
this may not be the case for more complicated systems. For example, at
higher aspect ratios, multiple stable solution branches 
may exist, and there is no guarantee that AE and SE will
arrive at the same solution. 
Some assumptions which inform the AE solution, such as the assumption
that the convection
intially occupies the same space as the evolved convection, may not hold in
studies of penetrative convection, despite the fact that similar methods have
long been used in those studies \cite{hurlburt&all1986}. Extensions to 
fully compressible convection in which there are two true thermodynamic
variables \cite{anders&brown2017} must contain a very careful treatment of
AE pressure adjustments, so as to avoid wave-launching pressure mismatches.
Our work here serves as a basis for determining if AE techniques are
effective in more complex studies of convection.

\begin{acknowledgments}
We thank Geoff Vasil, Daniel Lecoanet, 
and the two anonymous referees whose careful comments greatly 
improved the clarity and scientific content of this paper.
EHA acknowledges the support of the University of Colorado's George 
Ellery Hale Graduate Student Fellowship.
This work was additionally supported by  NASA LWS grant number NNX16AC92G.  
Computations were conducted 
with the support of the NASA High End Computing (HEC) Program through the NASA 
Advanced Supercomputing (NAS) Division at Ames Research Center on Pleiades
with allocation GID s1647.
\end{acknowledgments}

\appendix
\section{Table of Runs}
\label{appendix:run_table}
In Table \ref{table:run_parameters} we list key properties of all simulations
conducted in this work.  
\begin{table}
\caption{Simulation parameters. We report the supercriticality ($S$), Rayleigh number (Ra), 
and coefficient resolution (nz, nx, and ny are the number of coefficients in the
z, x, and y directions respectively).
Simulation run times required to reach convergence
are reported for the SE solutions ($t_{\text{SE}}$) and the AE solutions ($t_{\text{AE}}$).
The amount of time over which simulations measurements were taken in the evolved
state is listed ($t_{\text{avg}}$). All times are in freefall time units.  
The volume-averaged Nusselt number (Nu) of the
AE and SE solutions are shown.
In the upper part of the table, information pertaining to 2D runs is reported,
while information pertaining to 3D runs is in the lower part of the table.
}
\label{table:run_parameters}
\begin{center}
\begin{tabularx}{\textwidth}{ X X r | X X X | X X }
\hline																	
$S$	&	Ra	&	nz$\times$nx$\times$ny$\,\,\,\,\,\,\,$	&	$t_{\text{SE}}$	&	$t_{\text{AE}}$	&	$t_{\text{avg}}$	&	Nu$_{\text{SE}}$	&	Nu$_{\text{AE}}$	\\
\hline \hline \multicolumn{8}{c}{\vspace{-0.2cm}}\\
\multicolumn{8}{c}{\vspace{0.1cm}2D Runs} \\
\hline
$10^{1/3}$	    &	$2.79 \times 10^3$       &	32$\times$64$\,\,\,\,\,\,\,$&	$52.8$	&	$340$	&	100	&	1.46	&	1.46	\\
$10^{2/3}$	    &	$6.01 \times 10^3$       &	32$\times$64$\,\,\,\,\,\,\,$&	$77.6$	&	$282$	&	100	&	1.95	&	1.95	\\
$10^1$	        &	$1.30 \times 10^4$       &  	32$\times$64$\,\,\,\,\,\,\,$&	$114$	&	$265$	&	100	&	2.43	&	2.42	\\
$10^{1 + 1/3}$	&	$2.79 \times 10^4$       &	32$\times$64$\,\,\,\,\,\,\,$&	$167$	&	$251$	&	100	&	2.54	&	2.54	\\
$10^{1 + 2/3}$	&	$6.01 \times 10^4$       &	32$\times$64$\,\,\,\,\,\,\,$&	$245$	&	$245$	&	100	&	3.14	&	3.14	\\
$10^2$	        &	$1.30 \times 10^5$       &	64$\times$128$\,\,\,\,\,\,\,$&	$360$	&	$326$	&	100	&	3.8	&	3.8	\\
$10^{2 + 1/3}$	&	$2.79 \times 10^5$       &	64$\times$128$\,\,\,\,\,\,\,$&	$528$	&	$248$	&	100	&	4.71	&	4.71	\\
$10^{2 + 2/3}$	&	$6.01 \times 10^5$       &	64$\times$128$\,\,\,\,\,\,\,$&	$776$	&	$251$	&	100	&	5.5	&	5.5	\\
$10^3$	        &	$1.30 \times 10^6$       &	128$\times$256$\,\,\,\,\,\,\,$&	$1.14 \times 10^3$	&	$268$	&	200	&	6.4	&	6.33	\\
$10^{3 + 1/3}$	&	$2.79 \times 10^6$       &	128$\times$256$\,\,\,\,\,\,\,$&	$1.67 \times 10^3$	&	$247$	&	500	&	6.87	&	6.95	\\
$10^{3 + 2/3}$	&	$6.01 \times 10^6$       &	256$\times$512$\,\,\,\,\,\,\,$&	$2.45 \times 10^3$	&	$275$	&	500	&	7.54	&	7.59	\\
$10^4$	        &	$1.30 \times 10^7$       &	256$\times$512$\,\,\,\,\,\,\,$&	$3.60 \times 10^3$	&	$301$	&	500	&	8.83	&	8.83	\\
$10^{4 + 1/3}$	&	$2.79 \times 10^7$       &	256$\times$512$\,\,\,\,\,\,\,$&	$5.28 \times 10^3$	&	$317$	&	500	&	10.13	&	10.14	\\
$10^{4 + 2/3}$	&	$6.01 \times 10^7$       &	256$\times$512$\,\,\,\,\,\,\,$&	$7.76 \times 10^3$	&	$326$	&	500	&	11.65	&	11.69	\\
$10^5$	        &	$1.30 \times 10^8$       &	512$\times$1024$\,\,\,\,\,\,\,$&	$1.14 \times 10^4$	&	$411$	&	500	&	14.02	&	14.18	\\
$10^{5 + 1/3}$	&	$2.79 \times 10^8$       &	512$\times$1024$\,\,\,\,\,\,\,$&	$1.67 \times 10^4$	&	$391$	&	500	&	---	&	16.21	\\
$10^{5 + 2/3}$	&	$6.01 \times 10^8$       &	512$\times$1024$\,\,\,\,\,\,\,$&	$2.45 \times 10^4$	&	$453$	&	500	&	---	&	18.58	\\
$10^6$	        &	$1.30 \times 10^9$       &	1024$\times$2048$\,\,\,\,\,\,\,$	&	$3.60 \times 10^4$	&	$436$	&	500	&	---	&	22.13	\\
$10^7$	        &	$1.30 \times 10^{10}$	&	2048$\times$4096$\,\,\,\,\,\,\,$	&	$1.14 \times 10^5$	&	$183$	&	170	&	---	&	38.29	\\
\\ \hline \hline \multicolumn{8}{c}{\vspace{-0.2cm}}\\
\multicolumn{8}{c}{\vspace{0.1cm}3D Runs} \\
\hline
$10^1$          &	$1.30 \times 10^4$	&	32$\times$64$\times$64$\,\,\,\,\,\,\,$   &	$114$	&	$261$	&	100	&	2.42	&	2.42	\\
$10^2$          &	$1.30 \times 10^5$	&	64$\times$128$\times$128$\,\,\,\,\,\,\,$   &	$360$	&	$249$	&	100	&	3.97	&	4	\\
$10^3$          &	$1.30 \times 10^6$	&	128$\times$256$\times$256$\,\,\,\,\,\,\,$	&	$1.14 \times 10^3$	&	$243$	&	500	&	6.27	&	6.27	\\
$10^4$          &	$1.30 \times 10^7$	&	256$\times$512$\times$512$\,\,\,\,\,\,\,$	&	$3.60 \times 10^3$	&	$244$	&	500	&	9.92	&	9.88	\\
\hline																	
\end{tabularx}
\end{center}
\end{table}

\newpage
\section{Accelerated Evolution Recipe}
\label{appendix:recipe}
In order to achieve Accelerated Evolution (AE), we pause the Direct Numerical Simulation (DNS)
which is evolving the dynamics of convection and solve a 1D Boundary Value Problem (BVP)
consisting of Eqns.~(\ref{eqn:bouss_BVP_momentum}) \& (\ref{eqn:bouss_BVP_energy}).
After solving this BVP, we appropriately adjust the fields being evolved in the DNS
towards their evolved state, and then we continue running the now-evolved DNS.
The specific steps taken in completing the AE method are as follows:
\begin{enumerate}
\item Wait some time, $t_{\text{transient}}$, before beginning the AE process.
\item During the DNS, calculate time averages of the 1D vertical profiles of
F$_{\text{E}}$, F$_{\text{tot}}$, 
and $\angles{\bm{u} \times \bm{\omega}}_{x,y}$, updating them every timestep.  
To calculate these
averages, we use a trapezoidal-rule integration in time, and then divide by the
total time elapsed over which the average is taken. 
\item Pause the DNS once the averages are sufficiently converged. 
To ensure that an average is converged, at
least some time $t_{\text{min}}$ must have passed since the average was started to
ensure that the full range of convective dynamics are probed, and
the profiles must change by no more than $P$\% on a given timestep.
\item Construct $\xi$, F$_{\text{E, ev}}$, and $\angles{\bm{u} \times \bm{\omega}}_{x,y\text{, ev}}$,
as specified in Sec. \ref{sec:ae}
from the averaged profiles.
\item Solve the BVP for $\angles{T_1}_{x,y}$ and $\angles{\varpi}_{x,y}$ of the
evolved state.  Set the horizontal average of the current DNS thermodynamic fields
equal to the results of the BVP.
\item Multiply the velocity field, $\bm{u} = u\hat{x} + v\hat{y} + w\hat{z}$,
and the temperature fluctuations, $T - \angles{T}_{x,y}$,
by $\sqrt{\xi}$ in the DNS to properly reduce the convective flux.
\item Continue running the DNS.
\end{enumerate}
We refer to this process as an ``AE BVP solve.''

While the use of a single AE BVP solve rapidly advances the convecting state to
one that is closer to the relaxed state, we find that repeating this method 
multiple times is the best way to
ensure that the AE solution is truly converged. For all runs in 2D at $S < 10^5$, we
set $t_{\text{transient}} = 50$, completed an AE BVP solve
with $t_{\text{min}} = 30$ and $P = 0.1$, and then repeated the procedure.
For all 3D runs and 2D runs with $S \in [10^5, 10^6]$,
we did a first AE BVP solve with $t_{\text{transient}} = 20$,
$t_{\text{min}} = 20$, and $P = 1$ in order to quickly reach a near-
converged state and vastly increase our timestep size.  After this first solve, 
we completed two AE BVP solves, with $t_{\text{transient}} = 30$,
$t_{\text{min}} = 30$, 
and $P = 0.1$ to get very close to the solution (as in Fig.~\ref{fig:time_trace}c).
At very high $S = 10^7$, we ran two AE BVP solves with $t_{\text{min}} = 20$ and
$P = 1$. For the first solve, we set $t_{\text{transient}} = 20$, and for the
second we set $t_{\text{transient}} = 30$. We used fewer solves at this high
value of $S$ in part to reduce the computational expense of the run, and in
part because a third BVP generally did not greatly alter the solution
(as in Fig.~\ref{fig:time_trace}c, arrow 3). We wait 50 freefall times after
the final AE BVP solve of each run before beginning to take measurements.

In general, to use AE, a threshold, $f$, should be chosen. When the fractional
change of the mean temperature profile from an AE BVP solve becomes less than $f$,
the solution can be considered converged on its chosen solution branch.
In other words, once
\begin{equation}
\frac{|\angles{T}_{\text{DNS}} - \angles{T}_{\text{AE}}|}{|\angles{T}_{\text{DNS}}|} < f,
\end{equation}
the solution is converged. In this work, we chose our number of AE iterations
such that $f \approx 10^{-2}$.  In general, the user of AE could
set $f$ smaller, and in doing so reduce the separation between the AE and SE
solutions, which can be seen in e.g., Fig.~\ref{fig:temp_comparison}a\&b. However,
smaller values of $f$ require additional AE BVP solves, and likely require smaller values
of $P$, resulting in longer wait times while the horizontal averages are computed.


\begin{thebibliography}{30}%
\makeatletter
\providecommand \@ifxundefined [1]{%
 \@ifx{#1\undefined}
}%
\providecommand \@ifnum [1]{%
 \ifnum #1\expandafter \@firstoftwo
 \else \expandafter \@secondoftwo
 \fi
}%
\providecommand \@ifx [1]{%
 \ifx #1\expandafter \@firstoftwo
 \else \expandafter \@secondoftwo
 \fi
}%
\providecommand \natexlab [1]{#1}%
\providecommand \enquote  [1]{``#1''}%
\providecommand \bibnamefont  [1]{#1}%
\providecommand \bibfnamefont [1]{#1}%
\providecommand \citenamefont [1]{#1}%
\providecommand \href@noop [0]{\@secondoftwo}%
\providecommand \href [0]{\begingroup \@sanitize@url \@href}%
\providecommand \@href[1]{\@@startlink{#1}\@@href}%
\providecommand \@@href[1]{\endgroup#1\@@endlink}%
\providecommand \@sanitize@url [0]{\catcode `\\12\catcode `\$12\catcode
  `\&12\catcode `\#12\catcode `\^12\catcode `\_12\catcode `\%12\relax}%
\providecommand \@@startlink[1]{}%
\providecommand \@@endlink[0]{}%
\providecommand \url  [0]{\begingroup\@sanitize@url \@url }%
\providecommand \@url [1]{\endgroup\@href {#1}{\urlprefix }}%
\providecommand \urlprefix  [0]{URL }%
\providecommand \Eprint [0]{\href }%
\providecommand \doibase [0]{http://dx.doi.org/}%
\providecommand \selectlanguage [0]{\@gobble}%
\providecommand \bibinfo  [0]{\@secondoftwo}%
\providecommand \bibfield  [0]{\@secondoftwo}%
\providecommand \translation [1]{[#1]}%
\providecommand \BibitemOpen [0]{}%
\providecommand \bibitemStop [0]{}%
\providecommand \bibitemNoStop [0]{.\EOS\space}%
\providecommand \EOS [0]{\spacefactor3000\relax}%
\providecommand \BibitemShut  [1]{\csname bibitem#1\endcsname}%
\let\auto@bib@innerbib\@empty
\bibitem [{\citenamefont {{Brown}}\ \emph {et~al.}(2010)\citenamefont
  {{Brown}}, \citenamefont {{Browning}}, \citenamefont {{Brun}}, \citenamefont
  {{Miesch}},\ and\ \citenamefont {{Toomre}}}]{brown&all2010}%
  \BibitemOpen
  \bibfield  {author} {\bibinfo {author} {\bibfnamefont {B.~P.}\ \bibnamefont
  {{Brown}}}, \bibinfo {author} {\bibfnamefont {M.~K.}\ \bibnamefont
  {{Browning}}}, \bibinfo {author} {\bibfnamefont {A.~S.}\ \bibnamefont
  {{Brun}}}, \bibinfo {author} {\bibfnamefont {M.~S.}\ \bibnamefont
  {{Miesch}}}, \ and\ \bibinfo {author} {\bibfnamefont {J.}~\bibnamefont
  {{Toomre}}},\ }\bibfield  {title} {\enquote {\bibinfo {title} {{Persistent
  Magnetic Wreaths in a Rapidly Rotating Sun}},}\ }\href {\doibase
  10.1088/0004-637X/711/1/424} {\bibfield  {journal} {\bibinfo  {journal}
  {\apj}\ }\textbf {\bibinfo {volume} {711}},\ \bibinfo {pages} {424--438}
  (\bibinfo {year} {2010})},\ \Eprint {http://arxiv.org/abs/1011.2831}
  {arXiv:1011.2831 [astro-ph.SR]} \BibitemShut {NoStop}%
\bibitem [{\citenamefont {{Featherstone}}\ and\ \citenamefont
  {{Hindman}}(2016)}]{featherstone&hindman2016}%
  \BibitemOpen
  \bibfield  {author} {\bibinfo {author} {\bibfnamefont {N.~A.}\ \bibnamefont
  {{Featherstone}}}\ and\ \bibinfo {author} {\bibfnamefont {B.~W.}\
  \bibnamefont {{Hindman}}},\ }\bibfield  {title} {\enquote {\bibinfo {title}
  {{The Spectral Amplitude of Stellar Convection and Its Scaling in the
  High-Rayleigh-number Regime}},}\ }\href {\doibase 10.3847/0004-637X/818/1/32}
  {\bibfield  {journal} {\bibinfo  {journal} {\apj}\ }\textbf {\bibinfo
  {volume} {818}},\ \bibinfo {eid} {32} (\bibinfo {year} {2016})},\ \Eprint
  {http://arxiv.org/abs/1511.02396} {arXiv:1511.02396 [astro-ph.SR]}
  \BibitemShut {NoStop}%
\bibitem [{\citenamefont {{Viallet}}\ \emph {et~al.}(2011)\citenamefont
  {{Viallet}}, \citenamefont {{Baraffe}},\ and\ \citenamefont
  {{Walder}}}]{viallet&all2011}%
  \BibitemOpen
  \bibfield  {author} {\bibinfo {author} {\bibfnamefont {M.}~\bibnamefont
  {{Viallet}}}, \bibinfo {author} {\bibfnamefont {I.}~\bibnamefont
  {{Baraffe}}}, \ and\ \bibinfo {author} {\bibfnamefont {R.}~\bibnamefont
  {{Walder}}},\ }\bibfield  {title} {\enquote {\bibinfo {title} {{Towards a new
  generation of multi-dimensional stellar evolution models: development of an
  implicit hydrodynamic code}},}\ }\href {\doibase 10.1051/0004-6361/201016374}
  {\bibfield  {journal} {\bibinfo  {journal} {Astronomy \& Astrophysics}\
  }\textbf {\bibinfo {volume} {531}},\ \bibinfo {eid} {A86} (\bibinfo {year}
  {2011})},\ \Eprint {http://arxiv.org/abs/1103.1524} {arXiv:1103.1524
  [astro-ph.IM]} \BibitemShut {NoStop}%
\bibitem [{\citenamefont {{Viallet}}\ \emph {et~al.}(2013)\citenamefont
  {{Viallet}}, \citenamefont {{Baraffe}},\ and\ \citenamefont
  {{Walder}}}]{viallet&all2013}%
  \BibitemOpen
  \bibfield  {author} {\bibinfo {author} {\bibfnamefont {M.}~\bibnamefont
  {{Viallet}}}, \bibinfo {author} {\bibfnamefont {I.}~\bibnamefont
  {{Baraffe}}}, \ and\ \bibinfo {author} {\bibfnamefont {R.}~\bibnamefont
  {{Walder}}},\ }\bibfield  {title} {\enquote {\bibinfo {title} {{Comparison of
  different nonlinear solvers for 2D time-implicit stellar hydrodynamics}},}\
  }\href {\doibase 10.1051/0004-6361/201220725} {\bibfield  {journal} {\bibinfo
   {journal} {Astronomy \& Astrophysics}\ }\textbf {\bibinfo {volume} {555}},\
  \bibinfo {eid} {A81} (\bibinfo {year} {2013})},\ \Eprint
  {http://arxiv.org/abs/1305.6581} {arXiv:1305.6581 [astro-ph.SR]} \BibitemShut
  {NoStop}%
\bibitem [{\citenamefont {{Viallet}}\ \emph {et~al.}(2016)\citenamefont
  {{Viallet}}, \citenamefont {{Goffrey}}, \citenamefont {{Baraffe}},
  \citenamefont {{Folini}}, \citenamefont {{Geroux}}, \citenamefont {{Popov}},
  \citenamefont {{Pratt}},\ and\ \citenamefont {{Walder}}}]{viallet&all2016}%
  \BibitemOpen
  \bibfield  {author} {\bibinfo {author} {\bibfnamefont {M.}~\bibnamefont
  {{Viallet}}}, \bibinfo {author} {\bibfnamefont {T.}~\bibnamefont
  {{Goffrey}}}, \bibinfo {author} {\bibfnamefont {I.}~\bibnamefont
  {{Baraffe}}}, \bibinfo {author} {\bibfnamefont {D.}~\bibnamefont {{Folini}}},
  \bibinfo {author} {\bibfnamefont {C.}~\bibnamefont {{Geroux}}}, \bibinfo
  {author} {\bibfnamefont {M.~V.}\ \bibnamefont {{Popov}}}, \bibinfo {author}
  {\bibfnamefont {J.}~\bibnamefont {{Pratt}}}, \ and\ \bibinfo {author}
  {\bibfnamefont {R.}~\bibnamefont {{Walder}}},\ }\bibfield  {title} {\enquote
  {\bibinfo {title} {{A Jacobian-free Newton-Krylov method for time-implicit
  multidimensional hydrodynamics. Physics-based preconditioning for sound waves
  and thermal diffusion}},}\ }\href {\doibase 10.1051/0004-6361/201527339}
  {\bibfield  {journal} {\bibinfo  {journal} {Astronomy \& Astrophysics}\
  }\textbf {\bibinfo {volume} {586}},\ \bibinfo {eid} {A153} (\bibinfo {year}
  {2016})},\ \Eprint {http://arxiv.org/abs/1512.03662} {arXiv:1512.03662
  [astro-ph.IM]} \BibitemShut {NoStop}%
\bibitem [{\citenamefont {{Lecoanet}}\ \emph {et~al.}(2014)\citenamefont
  {{Lecoanet}}, \citenamefont {{Brown}}, \citenamefont {{Zweibel}},
  \citenamefont {{Burns}}, \citenamefont {{Oishi}},\ and\ \citenamefont
  {{Vasil}}}]{lecoanet&all2014}%
  \BibitemOpen
  \bibfield  {author} {\bibinfo {author} {\bibfnamefont {D.}~\bibnamefont
  {{Lecoanet}}}, \bibinfo {author} {\bibfnamefont {B.~P.}\ \bibnamefont
  {{Brown}}}, \bibinfo {author} {\bibfnamefont {E.~G.}\ \bibnamefont
  {{Zweibel}}}, \bibinfo {author} {\bibfnamefont {K.~J.}\ \bibnamefont
  {{Burns}}}, \bibinfo {author} {\bibfnamefont {J.~S.}\ \bibnamefont
  {{Oishi}}}, \ and\ \bibinfo {author} {\bibfnamefont {G.~M.}\ \bibnamefont
  {{Vasil}}},\ }\bibfield  {title} {\enquote {\bibinfo {title} {{Conduction in
  Low Mach Number Flows. I. Linear and Weakly Nonlinear Regimes}},}\ }\href
  {\doibase 10.1088/0004-637X/797/2/94} {\bibfield  {journal} {\bibinfo
  {journal} {\apj}\ }\textbf {\bibinfo {volume} {797}},\ \bibinfo {eid} {94}
  (\bibinfo {year} {2014})},\ \Eprint {http://arxiv.org/abs/1410.5424}
  {arXiv:1410.5424 [astro-ph.SR]} \BibitemShut {NoStop}%
\bibitem [{\citenamefont {{Anders}}\ and\ \citenamefont
  {{Brown}}(2017)}]{anders&brown2017}%
  \BibitemOpen
  \bibfield  {author} {\bibinfo {author} {\bibfnamefont {E.~H.}\ \bibnamefont
  {{Anders}}}\ and\ \bibinfo {author} {\bibfnamefont {B.~P.}\ \bibnamefont
  {{Brown}}},\ }\bibfield  {title} {\enquote {\bibinfo {title} {{Convective
  heat transport in stratified atmospheres at low and high Mach number}},}\
  }\href {\doibase 10.1103/PhysRevFluids.2.083501} {\bibfield  {journal}
  {\bibinfo  {journal} {Physical Review Fluids}\ }\textbf {\bibinfo {volume}
  {2}},\ \bibinfo {eid} {083501} (\bibinfo {year} {2017})},\ \Eprint
  {http://arxiv.org/abs/1611.06580} {arXiv:1611.06580 [physics.flu-dyn]}
  \BibitemShut {NoStop}%
\bibitem [{\citenamefont {{Bordwell}}\ \emph {et~al.}(2018)\citenamefont
  {{Bordwell}}, \citenamefont {{Brown}},\ and\ \citenamefont
  {{Oishi}}}]{bordwell&all2018}%
  \BibitemOpen
  \bibfield  {author} {\bibinfo {author} {\bibfnamefont {B.}~\bibnamefont
  {{Bordwell}}}, \bibinfo {author} {\bibfnamefont {B.~P.}\ \bibnamefont
  {{Brown}}}, \ and\ \bibinfo {author} {\bibfnamefont {J.~S.}\ \bibnamefont
  {{Oishi}}},\ }\bibfield  {title} {\enquote {\bibinfo {title} {{Convective
  Dynamics and Disequilibrium Chemistry in the Atmospheres of Giant Planets and
  Brown Dwarfs}},}\ }\href {\doibase 10.3847/1538-4357/aaa551} {\bibfield
  {journal} {\bibinfo  {journal} {\apj}\ }\textbf {\bibinfo {volume} {854}},\
  \bibinfo {eid} {8} (\bibinfo {year} {2018})},\ \Eprint
  {http://arxiv.org/abs/1802.03026} {arXiv:1802.03026 [astro-ph.EP]}
  \BibitemShut {NoStop}%
\bibitem [{\citenamefont {{Stix}}(2003)}]{stix2003}%
  \BibitemOpen
  \bibfield  {author} {\bibinfo {author} {\bibfnamefont {M.}~\bibnamefont
  {{Stix}}},\ }\bibfield  {title} {\enquote {\bibinfo {title} {{On the time
  scale of energy transport in the sun}},}\ }\href {\doibase
  10.1023/A:1022952621810} {\bibfield  {journal} {\bibinfo  {journal} {Solar
  Physics}\ }\textbf {\bibinfo {volume} {212}},\ \bibinfo {pages} {3--6}
  (\bibinfo {year} {2003})}\BibitemShut {NoStop}%
\bibitem [{\citenamefont {{Johnston}}\ and\ \citenamefont
  {{Doering}}(2009)}]{johnston&doering2009}%
  \BibitemOpen
  \bibfield  {author} {\bibinfo {author} {\bibfnamefont {H.}~\bibnamefont
  {{Johnston}}}\ and\ \bibinfo {author} {\bibfnamefont {C.~R.}\ \bibnamefont
  {{Doering}}},\ }\bibfield  {title} {\enquote {\bibinfo {title} {{Comparison
  of Turbulent Thermal Convection between Conditions of Constant Temperature
  and Constant Flux}},}\ }\href {\doibase 10.1103/PhysRevLett.102.064501}
  {\bibfield  {journal} {\bibinfo  {journal} {Phys. Rev. Lett.}\ }\textbf
  {\bibinfo {volume} {102}},\ \bibinfo {eid} {064501} (\bibinfo {year}
  {2009})},\ \Eprint {http://arxiv.org/abs/0811.0401} {arXiv:0811.0401
  [physics.flu-dyn]} \BibitemShut {NoStop}%
\bibitem [{\citenamefont {{Verzicco}}\ and\ \citenamefont
  {{Camussi}}(1997)}]{verzicco&camussi1997}%
  \BibitemOpen
  \bibfield  {author} {\bibinfo {author} {\bibfnamefont {R.}~\bibnamefont
  {{Verzicco}}}\ and\ \bibinfo {author} {\bibfnamefont {R.}~\bibnamefont
  {{Camussi}}},\ }\bibfield  {title} {\enquote {\bibinfo {title} {{Transitional
  regimes of low-Prandtl thermal convection in a cylindrical cell}},}\ }\href
  {\doibase 10.1063/1.869244} {\bibfield  {journal} {\bibinfo  {journal}
  {Physics of Fluids}\ }\textbf {\bibinfo {volume} {9}},\ \bibinfo {pages}
  {1287--1295} (\bibinfo {year} {1997})}\BibitemShut {NoStop}%
\bibitem [{\citenamefont {{Hurlburt}}\ \emph {et~al.}(1984)\citenamefont
  {{Hurlburt}}, \citenamefont {{Toomre}},\ and\ \citenamefont
  {{Massaguer}}}]{hurlburt&all1984}%
  \BibitemOpen
  \bibfield  {author} {\bibinfo {author} {\bibfnamefont {N.~E.}\ \bibnamefont
  {{Hurlburt}}}, \bibinfo {author} {\bibfnamefont {J.}~\bibnamefont
  {{Toomre}}}, \ and\ \bibinfo {author} {\bibfnamefont {J.~M.}\ \bibnamefont
  {{Massaguer}}},\ }\bibfield  {title} {\enquote {\bibinfo {title}
  {{Two-dimensional compressible convection extending over multiple scale
  heights}},}\ }\href {\doibase 10.1086/162235} {\bibfield  {journal} {\bibinfo
   {journal} {\apj}\ }\textbf {\bibinfo {volume} {282}},\ \bibinfo {pages}
  {557--573} (\bibinfo {year} {1984})}\BibitemShut {NoStop}%
\bibitem [{\citenamefont {{Couston}}\ \emph {et~al.}(2017)\citenamefont
  {{Couston}}, \citenamefont {{Lecoanet}}, \citenamefont {{Favier}},\ and\
  \citenamefont {{Le Bars}}}]{couston&all2017}%
  \BibitemOpen
  \bibfield  {author} {\bibinfo {author} {\bibfnamefont {L.-A.}\ \bibnamefont
  {{Couston}}}, \bibinfo {author} {\bibfnamefont {D.}~\bibnamefont
  {{Lecoanet}}}, \bibinfo {author} {\bibfnamefont {B.}~\bibnamefont
  {{Favier}}}, \ and\ \bibinfo {author} {\bibfnamefont {M.}~\bibnamefont {{Le
  Bars}}},\ }\bibfield  {title} {\enquote {\bibinfo {title} {{Dynamics of mixed
  convective-stably-stratified fluids}},}\ }\href {\doibase
  10.1103/PhysRevFluids.2.094804} {\bibfield  {journal} {\bibinfo  {journal}
  {Physical Review Fluids}\ }\textbf {\bibinfo {volume} {2}},\ \bibinfo {eid}
  {094804} (\bibinfo {year} {2017})},\ \Eprint
  {http://arxiv.org/abs/1709.06454} {arXiv:1709.06454 [physics.flu-dyn]}
  \BibitemShut {NoStop}%
\bibitem [{\citenamefont {{Brandenburg}}\ \emph {et~al.}(2005)\citenamefont
  {{Brandenburg}}, \citenamefont {{Chan}}, \citenamefont {{Nordlund}},\ and\
  \citenamefont {{Stein}}}]{brandenburg&all2005}%
  \BibitemOpen
  \bibfield  {author} {\bibinfo {author} {\bibfnamefont {A.}~\bibnamefont
  {{Brandenburg}}}, \bibinfo {author} {\bibfnamefont {K.~L.}\ \bibnamefont
  {{Chan}}}, \bibinfo {author} {\bibfnamefont {{\AA}.}~\bibnamefont
  {{Nordlund}}}, \ and\ \bibinfo {author} {\bibfnamefont {R.~F.}\ \bibnamefont
  {{Stein}}},\ }\bibfield  {title} {\enquote {\bibinfo {title} {{Effect of the
  radiative background flux in convection}},}\ }\href {\doibase
  10.1002/asna.200510411} {\bibfield  {journal} {\bibinfo  {journal}
  {Astronomische Nachrichten}\ }\textbf {\bibinfo {volume} {326}},\ \bibinfo
  {pages} {681--692} (\bibinfo {year} {2005})},\ \Eprint
  {http://arxiv.org/abs/astro-ph/0508404} {astro-ph/0508404} \BibitemShut
  {NoStop}%
\bibitem [{\citenamefont {{Stevens}}\ \emph {et~al.}(2011)\citenamefont
  {{Stevens}}, \citenamefont {{Lohse}},\ and\ \citenamefont
  {{Verzicco}}}]{stevens&all2011}%
  \BibitemOpen
  \bibfield  {author} {\bibinfo {author} {\bibfnamefont {R.~J.~A.~M.}\
  \bibnamefont {{Stevens}}}, \bibinfo {author} {\bibfnamefont {D.}~\bibnamefont
  {{Lohse}}}, \ and\ \bibinfo {author} {\bibfnamefont {R.}~\bibnamefont
  {{Verzicco}}},\ }\bibfield  {title} {\enquote {\bibinfo {title} {{Prandtl and
  Rayleigh number dependence of heat transport in high Rayleigh number thermal
  convection}},}\ }\href {\doibase 10.1017/jfm.2011.354} {\bibfield  {journal}
  {\bibinfo  {journal} {Journal of Fluid Mechanics}\ }\textbf {\bibinfo
  {volume} {688}},\ \bibinfo {pages} {31--43} (\bibinfo {year} {2011})},\
  \Eprint {http://arxiv.org/abs/1102.2307} {arXiv:1102.2307 [physics.flu-dyn]}
  \BibitemShut {NoStop}%
\bibitem [{\citenamefont {{Julien}}\ \emph {et~al.}(1998)\citenamefont
  {{Julien}}, \citenamefont {{Knobloch}},\ and\ \citenamefont
  {{Werne}}}]{julien&all1998}%
  \BibitemOpen
  \bibfield  {author} {\bibinfo {author} {\bibfnamefont {Keith}\ \bibnamefont
  {{Julien}}}, \bibinfo {author} {\bibfnamefont {Edgar}\ \bibnamefont
  {{Knobloch}}}, \ and\ \bibinfo {author} {\bibfnamefont {Joseph}\ \bibnamefont
  {{Werne}}},\ }\bibfield  {title} {\enquote {\bibinfo {title} {{A New Class of
  Equations for Rotationally Constrained Flows}},}\ }\href {\doibase
  10.1007/s001620050092} {\bibfield  {journal} {\bibinfo  {journal}
  {Theoretical and Computational Fluid Dynamics}\ }\textbf {\bibinfo {volume}
  {11}},\ \bibinfo {pages} {251--261} (\bibinfo {year} {1998})}\BibitemShut
  {NoStop}%
\bibitem [{\citenamefont {{Sprague}}\ \emph {et~al.}(2006)\citenamefont
  {{Sprague}}, \citenamefont {{Julien}}, \citenamefont {{Knobloch}},\ and\
  \citenamefont {{Werne}}}]{sprague&all2006}%
  \BibitemOpen
  \bibfield  {author} {\bibinfo {author} {\bibfnamefont {Michael}\ \bibnamefont
  {{Sprague}}}, \bibinfo {author} {\bibfnamefont {Keith}\ \bibnamefont
  {{Julien}}}, \bibinfo {author} {\bibfnamefont {Edgar}\ \bibnamefont
  {{Knobloch}}}, \ and\ \bibinfo {author} {\bibfnamefont {Joseph}\ \bibnamefont
  {{Werne}}},\ }\bibfield  {title} {\enquote {\bibinfo {title} {{Numerical
  simulation of an asymptotically reduced system for rotationally constrained
  convection}},}\ }\href {\doibase 10.1017/S0022112005008499} {\bibfield
  {journal} {\bibinfo  {journal} {Journal of Fluid Mechanics}\ }\textbf
  {\bibinfo {volume} {551}},\ \bibinfo {pages} {141--174} (\bibinfo {year}
  {2006})}\BibitemShut {NoStop}%
\bibitem [{\citenamefont {{Hurlburt}}\ \emph {et~al.}(1986)\citenamefont
  {{Hurlburt}}, \citenamefont {{Toomre}},\ and\ \citenamefont
  {{Massaguer}}}]{hurlburt&all1986}%
  \BibitemOpen
  \bibfield  {author} {\bibinfo {author} {\bibfnamefont {N.~E.}\ \bibnamefont
  {{Hurlburt}}}, \bibinfo {author} {\bibfnamefont {J.}~\bibnamefont
  {{Toomre}}}, \ and\ \bibinfo {author} {\bibfnamefont {J.~M.}\ \bibnamefont
  {{Massaguer}}},\ }\bibfield  {title} {\enquote {\bibinfo {title} {{Nonlinear
  compressible convection penetrating into stable layers and producing internal
  gravity waves}},}\ }\href {\doibase 10.1086/164796} {\bibfield  {journal}
  {\bibinfo  {journal} {\apj}\ }\textbf {\bibinfo {volume} {311}},\ \bibinfo
  {pages} {563--577} (\bibinfo {year} {1986})}\BibitemShut {NoStop}%
\bibitem [{\citenamefont {{Spiegel}}\ and\ \citenamefont
  {{Veronis}}(1960)}]{spiegel&veronis1960}%
  \BibitemOpen
  \bibfield  {author} {\bibinfo {author} {\bibfnamefont {E.~A.}\ \bibnamefont
  {{Spiegel}}}\ and\ \bibinfo {author} {\bibfnamefont {G.}~\bibnamefont
  {{Veronis}}},\ }\bibfield  {title} {\enquote {\bibinfo {title} {{On the
  Boussinesq Approximation for a Compressible Fluid.}}}\ }\href {\doibase
  10.1086/146849} {\bibfield  {journal} {\bibinfo  {journal} {\apj}\ }\textbf
  {\bibinfo {volume} {131}},\ \bibinfo {pages} {442} (\bibinfo {year}
  {1960})}\BibitemShut {NoStop}%
\bibitem [{\citenamefont {{Burns}}\ \emph {et~al.}(2016)\citenamefont
  {{Burns}}, \citenamefont {{Vasil}}, \citenamefont {{Oishi}}, \citenamefont
  {{Lecoanet}},\ and\ \citenamefont {{Brown}}}]{burns&all2016}%
  \BibitemOpen
  \bibfield  {author} {\bibinfo {author} {\bibfnamefont {K.}~\bibnamefont
  {{Burns}}}, \bibinfo {author} {\bibfnamefont {G.}~\bibnamefont {{Vasil}}},
  \bibinfo {author} {\bibfnamefont {J.}~\bibnamefont {{Oishi}}}, \bibinfo
  {author} {\bibfnamefont {D.}~\bibnamefont {{Lecoanet}}}, \ and\ \bibinfo
  {author} {\bibfnamefont {B.}~\bibnamefont {{Brown}}},\ }\href@noop {}
  {\enquote {\bibinfo {title} {{Dedalus: Flexible framework for spectrally
  solving differential equations}},}\ }\bibinfo {howpublished} {Astrophysics
  Source Code Library} (\bibinfo {year} {2016}),\ \Eprint
  {http://arxiv.org/abs/1603.015} {ascl:1603.015} \BibitemShut {NoStop}%
\bibitem [{\citenamefont {{Ascher}}\ \emph {et~al.}(1997)\citenamefont
  {{Ascher}}, \citenamefont {{Ruuth}},\ and\ \citenamefont
  {{Spiteri}}}]{ascher&all1997}%
  \BibitemOpen
  \bibfield  {author} {\bibinfo {author} {\bibfnamefont {U.~M.}\ \bibnamefont
  {{Ascher}}}, \bibinfo {author} {\bibfnamefont {S.~J.}\ \bibnamefont
  {{Ruuth}}}, \ and\ \bibinfo {author} {\bibfnamefont {R.~J.}\ \bibnamefont
  {{Spiteri}}},\ }\bibfield  {title} {\enquote {\bibinfo {title}
  {{Implicit-explicit Runge-Kutta methods for time-dependent partial
  differential equations}},}\ }\href@noop {} {\bibfield  {journal} {\bibinfo
  {journal} {Applied Numerical Mathematics}\ }\textbf {\bibinfo {volume}
  {25}},\ \bibinfo {pages} {151--167} (\bibinfo {year} {1997})}\BibitemShut
  {NoStop}%
\bibitem [{sup()}]{supp}%
  \BibitemOpen
  \href@noop {} {}\bibinfo {note} {{See Supplemental Material at [URL] for a
  .zip file of the python run scripts used to perform all simulations in this
  work. This code was run using the Dedalus mercurial repository
  (\url{https://bitbucket.org/dedalus-project/dedalus/src/default/}) at commit
  node ID dab6af2abaab03c059a4e9513f6a4d98320c2f02.}}\BibitemShut {Stop}%
\bibitem [{\citenamefont {Goluskin}\ \emph {et~al.}(2014)\citenamefont
  {Goluskin}, \citenamefont {Johnston}, \citenamefont {Flierl},\ and\
  \citenamefont {Spiegel}}]{goluskin&all2014}%
  \BibitemOpen
  \bibfield  {author} {\bibinfo {author} {\bibfnamefont {David}\ \bibnamefont
  {Goluskin}}, \bibinfo {author} {\bibfnamefont {Hans}\ \bibnamefont
  {Johnston}}, \bibinfo {author} {\bibfnamefont {Glenn~R.}\ \bibnamefont
  {Flierl}}, \ and\ \bibinfo {author} {\bibfnamefont {Edward~A.}\ \bibnamefont
  {Spiegel}},\ }\bibfield  {title} {\enquote {\bibinfo {title} {Convectively
  driven shear and decreased heat flux},}\ }\href {\doibase
  10.1017/jfm.2014.577} {\bibfield  {journal} {\bibinfo  {journal} {J. Fluid
  Mech.}\ }\textbf {\bibinfo {volume} {759}},\ \bibinfo {pages} {360--385}
  (\bibinfo {year} {2014})}\BibitemShut {NoStop}%
\bibitem [{\citenamefont {Goluskin}(2016)}]{goluskin2016}%
  \BibitemOpen
  \bibfield  {author} {\bibinfo {author} {\bibfnamefont {David}\ \bibnamefont
  {Goluskin}},\ }\href {\doibase 10.1007/978-3-319-23941-5} {\emph {\bibinfo
  {title} {Internally Heated Convection and Rayleigh-B{\'{e}}nard
  Convection}}}\ (\bibinfo  {publisher} {Springer International Publishing},\
  \bibinfo {year} {2016})\BibitemShut {NoStop}%
\bibitem [{\citenamefont {{Cattaneo}}\ \emph {et~al.}(1991)\citenamefont
  {{Cattaneo}}, \citenamefont {{Brummell}}, \citenamefont {{Toomre}},
  \citenamefont {{Malagoli}},\ and\ \citenamefont
  {{Hurlburt}}}]{cattaneo&all1991}%
  \BibitemOpen
  \bibfield  {author} {\bibinfo {author} {\bibfnamefont {F.}~\bibnamefont
  {{Cattaneo}}}, \bibinfo {author} {\bibfnamefont {N.~H.}\ \bibnamefont
  {{Brummell}}}, \bibinfo {author} {\bibfnamefont {J.}~\bibnamefont
  {{Toomre}}}, \bibinfo {author} {\bibfnamefont {A.}~\bibnamefont
  {{Malagoli}}}, \ and\ \bibinfo {author} {\bibfnamefont {N.~E.}\ \bibnamefont
  {{Hurlburt}}},\ }\bibfield  {title} {\enquote {\bibinfo {title} {{Turbulent
  compressible convection}},}\ }\href {\doibase 10.1086/169814} {\bibfield
  {journal} {\bibinfo  {journal} {\apj}\ }\textbf {\bibinfo {volume} {370}},\
  \bibinfo {pages} {282--294} (\bibinfo {year} {1991})}\BibitemShut {NoStop}%
\bibitem [{\citenamefont {{Korre}}\ \emph {et~al.}(2017)\citenamefont
  {{Korre}}, \citenamefont {{Brummell}},\ and\ \citenamefont
  {{Garaud}}}]{korre&all2017}%
  \BibitemOpen
  \bibfield  {author} {\bibinfo {author} {\bibfnamefont {L.}~\bibnamefont
  {{Korre}}}, \bibinfo {author} {\bibfnamefont {N.}~\bibnamefont {{Brummell}}},
  \ and\ \bibinfo {author} {\bibfnamefont {P.}~\bibnamefont {{Garaud}}},\
  }\bibfield  {title} {\enquote {\bibinfo {title} {{Weakly non-Boussinesq
  convection in a gaseous spherical shell}},}\ }\href {\doibase
  10.1103/PhysRevE.96.033104} {\bibfield  {journal} {\bibinfo  {journal}
  {\pre}\ }\textbf {\bibinfo {volume} {96}},\ \bibinfo {eid} {033104} (\bibinfo
  {year} {2017})},\ \Eprint {http://arxiv.org/abs/1704.00817} {arXiv:1704.00817
  [physics.flu-dyn]} \BibitemShut {NoStop}%
\bibitem [{\citenamefont {{Stevens}}\ \emph {et~al.}(2010)\citenamefont
  {{Stevens}}, \citenamefont {{Verzicco}},\ and\ \citenamefont
  {{Lohse}}}]{stevens&all2010}%
  \BibitemOpen
  \bibfield  {author} {\bibinfo {author} {\bibfnamefont {Richard J.~A.~M.}\
  \bibnamefont {{Stevens}}}, \bibinfo {author} {\bibfnamefont {Roberto}\
  \bibnamefont {{Verzicco}}}, \ and\ \bibinfo {author} {\bibfnamefont {Detlef}\
  \bibnamefont {{Lohse}}},\ }\bibfield  {title} {\enquote {\bibinfo {title}
  {{Radial boundary layer structure and Nusselt number in Rayleigh-B{\'e}nard
  convection}},}\ }\href {\doibase 10.1017/S0022112009992461} {\bibfield
  {journal} {\bibinfo  {journal} {Journal of Fluid Mechanics}\ }\textbf
  {\bibinfo {volume} {643}},\ \bibinfo {pages} {495--507} (\bibinfo {year}
  {2010})}\BibitemShut {NoStop}%
\bibitem [{\citenamefont {{Otero}}\ \emph {et~al.}(2002)\citenamefont
  {{Otero}}, \citenamefont {{Wittenberg}}, \citenamefont {{Worthing}},\ and\
  \citenamefont {{Doering}}}]{otero&all2002}%
  \BibitemOpen
  \bibfield  {author} {\bibinfo {author} {\bibfnamefont {J.}~\bibnamefont
  {{Otero}}}, \bibinfo {author} {\bibfnamefont {R.~W.}\ \bibnamefont
  {{Wittenberg}}}, \bibinfo {author} {\bibfnamefont {R.~A.}\ \bibnamefont
  {{Worthing}}}, \ and\ \bibinfo {author} {\bibfnamefont {C.~R.}\ \bibnamefont
  {{Doering}}},\ }\bibfield  {title} {\enquote {\bibinfo {title} {{Bounds on
  Rayleigh B{\'e}nard convection with an imposed heat flux}},}\ }\href
  {\doibase 10.1017/S0022112002002410} {\bibfield  {journal} {\bibinfo
  {journal} {J. Fluid Mech.}\ }\textbf {\bibinfo {volume} {473}},\ \bibinfo
  {pages} {191--199} (\bibinfo {year} {2002})}\BibitemShut {NoStop}%
\bibitem [{\citenamefont {{Wall}}\ and\ \citenamefont
  {{Jenkins}}(2012)}]{wall&jenkins2012}%
  \BibitemOpen
  \bibfield  {author} {\bibinfo {author} {\bibfnamefont {J.~V.}\ \bibnamefont
  {{Wall}}}\ and\ \bibinfo {author} {\bibfnamefont {C.~R.}\ \bibnamefont
  {{Jenkins}}},\ }\href@noop {} {\emph {\bibinfo {title} {Practical Statistics
  for Astronomers, by J.~V.~Wall , C.~R.~Jenkins, Cambridge, UK: Cambridge
  University Press, 2012}}}\ (\bibinfo {year} {2012})\BibitemShut {NoStop}%
\bibitem [{\citenamefont {{K{\"a}pyl{\"a}}}\ \emph {et~al.}(2017)\citenamefont
  {{K{\"a}pyl{\"a}}}, \citenamefont {{Rheinhardt}}, \citenamefont
  {{Brandenburg}}, \citenamefont {{Arlt}}, \citenamefont {{K{\"a}pyl{\"a}}},
  \citenamefont {{Lagg}}, \citenamefont {{Olspert}},\ and\ \citenamefont
  {{Warnecke}}}]{kapyla&all2017}%
  \BibitemOpen
  \bibfield  {author} {\bibinfo {author} {\bibfnamefont {P.~J.}\ \bibnamefont
  {{K{\"a}pyl{\"a}}}}, \bibinfo {author} {\bibfnamefont {M.}~\bibnamefont
  {{Rheinhardt}}}, \bibinfo {author} {\bibfnamefont {A.}~\bibnamefont
  {{Brandenburg}}}, \bibinfo {author} {\bibfnamefont {R.}~\bibnamefont
  {{Arlt}}}, \bibinfo {author} {\bibfnamefont {M.~J.}\ \bibnamefont
  {{K{\"a}pyl{\"a}}}}, \bibinfo {author} {\bibfnamefont {A.}~\bibnamefont
  {{Lagg}}}, \bibinfo {author} {\bibfnamefont {N.}~\bibnamefont {{Olspert}}}, \
  and\ \bibinfo {author} {\bibfnamefont {J.}~\bibnamefont {{Warnecke}}},\
  }\bibfield  {title} {\enquote {\bibinfo {title} {{Extended Subadiabatic Layer
  in Simulations of Overshooting Convection}},}\ }\href {\doibase
  10.3847/2041-8213/aa83ab} {\bibfield  {journal} {\bibinfo  {journal}
  {Astrophys. J. Lett.}\ }\textbf {\bibinfo {volume} {845}},\ \bibinfo {eid}
  {L23} (\bibinfo {year} {2017})},\ \Eprint {http://arxiv.org/abs/1703.06845}
  {arXiv:1703.06845 [astro-ph.SR]} \BibitemShut {NoStop}%
\end{thebibliography}
\end{document}